\documentclass[12pt]{article}

\usepackage{amsmath}
\usepackage{amssymb}
\usepackage{graphicx}
\usepackage{axodraw}
\usepackage{subfigure}
\usepackage{url}
\usepackage[dvips,usenames]{color}

\begin{document}
\baselineskip 0.6cm

\def\simgt{\mathrel{\lower2.5pt\vbox{\lineskip=0pt\baselineskip=0pt
           \hbox{$>$}\hbox{$\sim$}}}}
\def\simlt{\mathrel{\lower2.5pt\vbox{\lineskip=0pt\baselineskip=0pt
           \hbox{$<$}\hbox{$\sim$}}}}
\def\simprop{\mathrel{\lower3.0pt\vbox{\lineskip=1.0pt\baselineskip=0pt
             \hbox{$\propto$}\hbox{$\sim$}}}}
\def\lg{\mathrel{\lower2.5pt\vbox{\lineskip=0pt\baselineskip=0pt
           \hbox{$<$}\hbox{$>$}}}}
\def\lrpartial{\stackrel{\leftrightarrow}{\partial}}
\def\lrD{\stackrel{\leftrightarrow}{\cal D}}
\def\rmax{{\rm max}}
\def\rmin{{\rm min}}

\begin{titlepage}

\begin{flushright}
MIT-CTP-4355 \\
UCB-PTH-12/05 \\

\end{flushright}

\vskip 1.0cm

\begin{center}
{\Large \bf What can the observation of nonzero curvature tell us?}

\vskip 0.6cm

{\large Alan H. Guth$^a$ and Yasunori Nomura$^b$}

\vskip 0.3cm

$^a$ {\it Center for Theoretical Physics, Laboratory for Nuclear Science, 
     and Department of Physics, \\
     Massachusetts Institute of Technology, Cambridge, MA 02139, USA} \\
$^b$ {\it Berkeley Center for Theoretical Physics, Department of Physics, \\
     and Theoretical Physics Group, Lawrence Berkeley National Laboratory, \\
     University of California, Berkeley, CA 94720, USA} \\

\vskip 0.5cm

\abstract{The eternally inflating multiverse provides a consistent 
 framework to understand coincidences and fine-tuning in the universe. 
 As such, it provides the possibility of finding another coincidence:\ 
 if the amount of slow-roll inflation in our past was only slightly 
 more than the anthropic threshold, then spatial curvature might be 
 measurable.  We study this issue in detail, particularly focusing 
 on the question:\ ``If future observations reveal nonzero curvature, 
 what can we conclude?''  We find that whether an observable signal 
 arises or not depends crucially on three issues:\ the cosmic history 
 just before the observable inflation, the measure adopted to define 
 probabilities in the eternally inflating spacetime, and the sign and 
 strength of the correlation between the tunneling and slow-roll parts 
 of the potential.  We find that if future measurements find positive 
 curvature at the level $\Omega_k \simlt -10^{-4}$, then the framework 
 of the eternally inflating multiverse, as currently understood, is 
 excluded with high significance.  If the measurements instead reveal 
 negative curvature at the level $\Omega_k \simgt 10^{-4}$, then we 
 can conclude that (1) diffusive (new or chaotic type) eternal inflation 
 did not occur in our immediate past; (2) our pocket universe was born 
 by a bubble nucleation; (3) the probability measure does not reward 
 volume increase; and (4) the origin of the observed slow-roll inflation 
 is an accidental feature of the potential, presumably selected by 
 anthropic conditions, and not due to a theoretical mechanism ensuring 
 the flatness of the potential. Discovery of $\Omega_k \simgt 10^{-4}$ 
 would also give us nontrivial information about the correlation 
 between the tunneling and slow-roll parts of the potential; for 
 example, a strong correlation favoring large $N$ would be ruled 
 out in certain measures.  We also address the question of whether 
 the current constraint on $\Omega_k$ is consistent with multiverse 
 expectations; we find the answer to be yes, except that current 
 observations, for many choices of measure, rule out the possibility 
 of strong correlations in the potential which favor small values 
 of $N$.  In the course of this work we were led to consider vacuum 
 decay branching ratios, and found that it is more likely than one 
 might guess that the decays are dominated by a single channel. 
 Planned future measurements of spatial curvature provide a valuable 
 opportunity to explore the structure of the multiverse as well as 
 the cosmic history just before the observable inflation.}

\end{center}
\end{titlepage}

\section{Introduction}
\label{sec:intro}

Evidence for cosmic inflation in the early history of our universe 
is mounting.  In addition to the original motivation of explaining 
flatness and homogeneity of the observable universe~\cite{Guth:1980zm}, 
we now have precision data from the cosmic microwave background 
(CMB)~\cite{Komatsu:2008hk} that is in beautiful agreement with the 
predictions of the simplest inflationary models~\cite{Hawking:1982cz}. 
The details of this cosmic inflation, however, remain very uncertain. 
We do not know its energy scale, its duration, or the circumstances 
that led to its onset.

In the last decade, we have been learning that many of the structures of our 
own universe may be understood as a result of environmental, or anthropic, 
selection in the multiverse~\cite{Hogan:1999wh}.  The most successful outcome 
of this picture was the prediction of a nonzero cosmological constant, made 
already in the 1980's~\cite{Weinberg:1987dv} and confirmed in 1998 by the 
discovery of an accelerating expansion of the universe~\cite{Riess:1998cb}. 
The picture of the multiverse is motivated theoretically by eternal 
inflation~\cite{Guth:1982pn,Vilenkin:1983xq,Linde:1986fd} and the 
landscape~\cite{Bousso:2000xa} of string theory, which together provide 
a consistent framework for explaining the nonzero cosmological constant 
and other examples of fine-tuning in the universe.  The onset of cosmic 
inflation itself can perhaps be understood in the same way:\ since 
excessive curvature suppresses structure formation~\cite{Vilenkin:1996ar}, 
it is possible that we are living in the aftermath of an era of inflation 
because otherwise intelligent observers would not have evolved.

An interesting consequence of this picture is that the observable era 
of inflation---i.e., the last $N \approx (40~\mbox{--}~60)$ $e$-folds 
of inflation, which are probed by the density perturbations in the CMB 
and in the matter distribution---may have been ``just so.''  That is, 
the number of $e$-folds of the slow-roll inflation may have been very 
close to the minimal number needed to ensure the flatness required for 
the evolution of life.  Such a coincidence would seem unlikely in a more 
conventional picture, in which the flatness of the inflaton potential 
might be ensured, for example, by some approximate symmetry.  But in 
the context of the multiverse, such a coincidence is very plausible. 
This leads to a number of potentially observable signatures, especially 
in structures at large scales, including nonzero curvature of the 
universe~\cite{Garriga:1998px,Freivogel:2005vv}.  Studies along these 
lines have been performed, e.g., in Refs.~\cite{Tegmark:2004qd}.

Whether an observable signal actually arises or not, however, depends 
on at least three issues: 1) What was the cosmic history just before 
the observable era of inflation; 2) What probability ``measure'' is 
adopted to define probabilities in the eternally inflating spacetime, 
where anything that can happen will happen an infinite number of times; 
and 3) In tunneling transitions from one vacuum to another, how strong 
are the correlations between the tunneling rate and the properties 
of any slow-roll inflation that might follow the tunneling?  In this 
paper, we explore these issues, focusing on the question:\ ``If future 
observations reveal nonzero curvature, what can we conclude?''  We take 
a bottom-up approach---we consider a variety of possibilities for the 
pre-inflationary history and the multiverse measure, which we think 
are reasonably exhaustive, and we consider both strong and weak tunneling 
rate correlations.  For the pre-inflationary history, we consider four 
different classes of models, characterized by the behavior of the inflaton 
field prior to the observable era of inflation.  For the multiverse 
measure, we consider various geometric cutoff measures~\cite{Guth:2000ka} 
as well as the recently proposed quantum measure~\cite{Nomura:2011dt}, 
in which the probability is given by the quantum-mechanical Born rule 
applied to the multiverse state.  We will see that the observation of 
curvature beyond the level of $\sim 10^{-4}$ can either exclude the 
multiverse framework itself (if it is positive) or exclude certain 
pre-inflationary histories {\it and} classes of probability measures 
(if it is negative), as well as constrain the nature and degree of 
correlation between the tunneling rate for a transition and the ensuing 
slow-roll inflation.

In the next section, we carefully define the framework of our analysis. 
We begin by classifying possible pre-inflationary histories, and then 
we discuss probability measures.  Section~\ref{sec:f-N} provides the 
actual analysis.  The meaning of the probability distribution for 
curvature in the context of bubble universes is also elucidated 
there.  We analyze all the possible scenarios for the pre-inflationary 
histories as well as the probability measures.  Our result for the 
probability distribution for curvature (in the negative case) will 
be presented in Section~\ref{sec:expectation}.  We finally conclude 
in Section~\ref{sec:concl}, summarizing what we can learn from a future 
observation of nonzero curvature of the universe.  One appendix discusses 
the effect of volume increase in the quantum measure, and a second 
appendix discusses the possibility that vacuum decays might be dominated 
by a single channel.

While completing this paper we received Ref.~\cite{Kleban:2012ph}, by 
Kleban and Schillo, which also discusses the issue of spatial curvature 
and the cosmic history before the observable inflation.  Our conclusions 
about it are consistent with theirs.  In fact our treatment of scenario 
(iv) in Section~\ref{subsec:pre-inflation} is based on private 
communication with Kleban~\cite{Kleban}.

\section{Framework}
\label{sec:curvature}

The observable era of early-universe inflation---i.e., the last 
$N \approx (40~\mbox{--}~60)$ $e$-folds of inflation---was the period 
during which currently observable scales went outside the Hubble horizon.%
\footnote{We use the phrase ``Hubble horizon'' to denote the distance 
 scale $H^{-1}$, where $H$ is the Hubble parameter, although the actual 
 causal horizon is vastly larger.}
Cosmic history  before this era, however, can leave its imprint on the 
present-day curvature contribution, $\Omega_k \equiv 1 - \Omega_0$. 
The expected amount of curvature depends strongly on the cosmic history 
just before the observable inflation, the measure used to define 
probabilities in the eternally inflating spacetime, and the nature and 
degree of correlation between vacuum transition tunneling rates and the 
ensuing slow-roll inflation.  In this section, we consider a variety 
of assumptions on the first two issues, establishing a framework for 
the analyses in later sections.  The issue of the correlation between 
tunneling and slow-roll will be discussed in Section~\ref{subsec:bubble}, 
where it becomes relevant.

\subsection{History (just) before the observable inflation}
\label{subsec:pre-inflation}

Since the observable inflation occurred with energy densities much 
smaller than the Planckian density, the cosmic history just before it 
must be describable using (semi-)classical gravity.  Here we consider 
four scenarios, which we think cover most of the realistic possibilities:
\begin{itemize}
\item[(i)]
{\it Eternal New Inflation} --- By tracing history back in time, the 
inflaton field $\varphi$ reaches a local maximum of the potential, 
with an energy density significantly smaller than the Planck scale, 
i.e. $V_0 \ll M_{\rm Pl}^4 \equiv (8 \pi G_N)^{-2}$.  Denoting this 
point as $\varphi = 0$ (which may be a saddle point in multi-dimensional 
field space), the ``initial conditions'' are given by $\varphi \approx 
\dot{\varphi} \approx 0$.  The question of how these conditions arose 
need not concern us here, as the results for $\Omega_k$ are insensitive 
to how these conditions were prepared.  The dynamics beginning 
with these initial conditions is described by eternal inflation 
at $\varphi \approx 0$~\cite{Vilenkin:1983xq}, followed by slow-roll 
inflation occurring near the potential minimum that corresponds to 
our vacuum.
\item[(ii)]
{\it Eternal Chaotic Inflation} --- By tracing history back in time, 
the inflaton field climbs up a hill in the potential energy diagram 
to the point where the quantum fluctuation in the field $\varDelta 
\varphi_{\rm qu} \approx H/2\pi$ (averaged over a Hubble volume during 
a Hubble time interval) becomes so important that the global structure 
of spacetime is determined by $\varDelta \varphi_{\rm qu}$, rather than 
by classical evolution of the field.  Here, $H = (V/3M_{\rm Pl}^2)^{1/2}$ 
is the Hubble parameter.  The transition point for quantum-fluctuation 
dominance is at a super-Planckian field value $\varphi_* \gg M_{\rm Pl}$; 
for example, for $V = \frac{1}{2} m^2 \varphi^2$ it is at $\varphi_* 
\approx M_{\rm Pl}^{3/2}/m^{1/2}$, and for $V = \frac{1}{4!} \lambda 
\varphi^4$ it is at $\varphi_* \approx M_{\rm Pl}/\lambda^{1/6}$, 
where $m \ll M_{\rm Pl}$ (or $\lambda \ll 1$) to reproduce the 
observed magnitude of density perturbations.  The cosmic history 
before our big-bang universe is then described  by eternal chaotic 
inflation~\cite{Linde:1986fd} followed by slow-roll chaotic inflation.
\item[(iii)]
{\it Eternal Old Inflation} --- By tracing history back in time, we hit 
a quantum tunneling event before entering into an eternally inflating 
epoch.  Our pocket universe then arose directly from a bubble nucleation 
process~\cite{Coleman:1980aw}, presumably occurring in an eternally 
inflating region in which the inflaton field was in some local minimum 
of the potential~\cite{Guth:1982pn}.  The bubble nucleation is followed 
by a brief curvature-dominated epoch, followed by non-eternal slow-roll 
inflation~\cite{Gott:1982zf}.  Slow-roll inflation begins when the 
vacuum energy starts to dominate over the curvature and kinetic 
energies in determining the evolution of the bubble universe.
\item[(iv)]
{\it A Prior Episode of Inflation} --- By tracing history back in time, 
the dynamics of another scalar field becomes important.  This is the 
case, for example, in double-inflation~\cite{Silk:1986vc} or in hybrid 
inflation~\cite{Linde:1993cn} if the waterfall field is fully responsible 
for the observable inflation.  An interesting feature of this scenario 
is that the density fluctuation spectrum shows a sharp spike at 
the scale corresponding to the connection of the two inflationary 
periods~\cite{Randall:1995dj}.  The resulting perturbation in 
$\Omega_k$ can be either positive or negative, which is determined 
only stochastically.
\end{itemize}
For each of the four cases above, we can estimate the probability 
distribution for $\Omega_k$ under various assumptions about the 
probability measure, the a priori probability distribution for parameters 
in the inflaton potential, and (when relevant) the initial conditions 
after the tunneling event.

\subsection{Measures in eternal inflation}
\label{subsec:measures}

In an eternally inflating multiverse, anything that can happen will 
happen infinitely many times.  This implies, among others, the following 
two statements.  First, to define the relative likelihood of different 
types of events, we need to regularize the infinities.  Second, any 
prediction in the multiverse will necessarily be statistical.  Here 
we consider the first of these statements, leaving the second to the 
next subsection.

Regularizing infinities in the multiverse has been an extensive area 
of research~\cite{Guth:2000ka}.  There have been many proposals for 
``measures'' that provide required regularizations, and thus prescriptions 
for making predictions.  Traditionally, these measures have been defined 
using ``global'' or ``local'' geometric cutoffs (although this division is 
not always meaningful, since the same measure can often be formulated using 
either a global or local description~\cite{Bousso:2008hz,Bousso:2009dm}). 
Global-cutoff measures propose that relative probabilities can be 
determined by the ratio of the number of events that occur prior to 
a specified ``equal-time'' hypersurface, usually in the limit as the 
hypersurface is chosen at an arbitrarily late time.  Depending on the 
choice of hypersurfaces, different measures can be obtained.  Local-cutoff 
measures, on the other hand, count events inside a finite neighborhood 
of a single timelike geodesic, and probabilities are computed after 
certain averaging procedures.  Different measures correspond to 
different choices for the neighborhood.

More recently, a framework for the eternally inflating multiverse has been 
proposed which does not rely on a geometric cutoff~\cite{Nomura:2011dt}. 
In this framework, the entire multiverse is a single quantum state 
as described {\it from a single reference frame}.  It is in general 
a superposition of many quantum states corresponding to well-defined 
semi-classical geometries, each of which is defined {\it only in and 
on} the apparent horizon. (This restriction on spacetime, dictated by 
the principles of quantum mechanics, provides the required regularization.) 
The well-defined probabilities are then given by the simple Born rule 
extended to the whole spacetime.  This framework allows us to use the 
same probability formula for questions regarding global properties of 
the universe and outcomes of particular experiments, thus providing 
complete unification of the eternally inflating multiverse and the 
many-worlds interpretation of quantum mechanics.

Is there a general classification scheme that accommodates all these 
measures and is relevant for our present purpose of discussing the curvature 
contribution to the universe?  A useful classification is obtained by 
considering how the measure does or does not reward the exponential 
increase in volume that characterizes inflationary models.  Here we 
discuss the following three classes, where examples of each appear 
in the literature:
\begin{itemize}
\item[(I)]
{\it Measures rewarding any volume increase} --- These measures reward 
any volume increase in the evolution of the multiverse.  The simplest 
example is the so-called proper-time cutoff measure~\cite{Linde:1993nz}, 
which defines probabilities in the global picture using hypersurfaces 
of equal proper time, obtained through the congruence of geodesics 
orthogonal to some arbitrary initial hypersurface.  This class of 
measures, however, suffers from various difficulties.  The most serious 
one is probably the youngness paradox~\cite{Guth:2000ka-2}:\ because 
of the rapid expansion of spatial volume in the eternally inflating 
region, the population of pocket universes is extremely youth-dominated. 
The probability of observing a universe that is old like ours (with 
$T_{\rm CMB} \simeq 2.7~{\rm K}$) becomes vanishingly small.  Since 
this essentially excludes observationally the class of measures 
described here, we will not consider it further.
\item[(II)]
{\it Measures rewarding volume increase only in the slow-roll 
regime} --- In these measures the volume increase during the eternally 
inflating regime is not rewarded, so the youngness paradox does not 
arise.  To model the behavior of these measures, suppose that the 
probability density for the {\it onset} of an episode of inflation 
of $N$ $e$-folds is given by some function $f(N)$.  That is, $f(N)\,dN$ 
is the probability that the number of inflationary $e$-folds that 
will follow a randomly selected onset of inflation will lie between 
$N$ and $N+dN$.  $f(N)$ would in principle be determined by the 
probability distributions for inflaton potential parameters and 
for the inflaton field in the multiverse.  While we do not know enough 
to compute $f(N)$, we will argue later that we can estimate its behavior 
under a variety of assumptions.  Once inflation begins, the volume of 
the inflated region is multiplied by $e^{3N}$, so the probability density 
$P(N)$ of finding oneself in a region that has undergone $N$ $e$-folds 
of slow-roll inflation can be written as
\begin{equation}
  P(N) \sim f(N) e^{3N}\, ,
 \label{eq:P-N_class-2}
\end{equation}
where $e^{3N}$ is the dominant factor.  While the class 
of measures considered here has issues that need to be 
addressed~\cite{Page:2006dt,Feldstein:2005bm}, it is not clear if 
these measures are excluded~\cite{Vilenkin:2006qg,Linde:2005yw}. 
We therefore keep these measures in our consideration.  An important 
example of this class of measures is given by the so-called 
pocket-based measure~\cite{Garriga:2005av}.
\item[(III)]
{\it Measures not rewarding volume increase} --- These measures 
do not reward volume increase due to any form of inflation.  Naively, 
this may sound rather counter-intuitive:\ how can a larger spatial 
volume avoid giving more observers, leading to a larger weight? 
This picture, however, can arise naturally in several different ways. 
For example, we could count events along a geodesic randomly chosen 
on an initial spacelike hypersurface, we could measure spacetime 
according to its comoving volume, or we could use a global time 
cutoff based on the total amount of expansion (i.e., scale-factor 
time).  The probability distribution for finding oneself in a region 
that has undergone $N$ $e$-folds of slow-roll inflation is then simply
\begin{equation}
  P(N) \sim f(N)\, .
\label{eq:P-N_class-3}
\end{equation}
The fact that volume increase is not rewarded in the final probability 
distribution makes it rather easy to avoid the problems encountered by 
measures of type (I) and (II).  Two examples of geometric cutoff measures 
in this class are the causal patch measure~\cite{Bousso:2006ev,Bousso:2009dm} 
and the scale-factor cutoff measure~\cite{DeSimone:2008bq}.  The recently 
proposed quantum-mechanical measure~\cite{Nomura:2011dt} also falls in 
this class, as discussed in Appendix~\ref{app:quantum}.
\end{itemize}

Equations~(\ref{eq:P-N_class-2}) and (\ref{eq:P-N_class-3}) can be 
summarized by writing
\begin{equation}
  P(N) \sim f(N) M_m(N)\, ,
\label{eq:P-N}
\end{equation}
where the dependence on the measure $m$ is described by the factor 
$M_m(N)$.  In this paper we are assuming that the measure is adequately 
described by specifying that it belongs to class (II) or class (III) 
above, so
\begin{equation}
  M_m(N) \approx \left\{ 
    \begin{array}{ll}
      e^{3N} & \hbox{ if\,\, $m \in {\rm (II)}$}\\
      1 & \hbox{ if\,\, $m \in {\rm (III)}$}\, .
  \end{array} \right.
\label{eq:P-N-M}
\end{equation}
It is, in principle, possible to consider hybrids of these classes. 
For example, in the stationary measure of Ref.~\cite{Linde:2007nm} 
features of both (I) and (II) coexist.  We will also comment on these 
hybrid possibilities when we discuss the probability distribution 
of $\Omega_k$ later.

\subsection{Probability distributions for current and future measurements}
\label{subsec:probability}

In order to discuss implications of a future measurement of curvature by 
our civilization, we can study the multiverse probability distribution for 
$\Omega_k$ as a conditional probability density, given the set of observed 
values of the physical parameters $\{ Q_1, Q_2, \ldots \}$ that have 
already been measured.  These parameters $\{Q_i\}$ include cosmological 
parameters such as the primordial density fluctuation amplitude 
$\delta\rho/\rho$, the scalar spectral index $n_s$, and the vacuum 
energy density $\rho_\Lambda$, as well as particle physics parameters 
such as the electron mass $m_e$, the proton mass $m_p$, the fine 
structure constant $\alpha$, etc.  The conditional probability density 
$f_{\rm cond}(\Omega_k \vert \{Q_i = Q_{i,\rm obs}\})$ is proportional 
to the full probability density function $f(\Omega_k, \{Q_i\})$ evaluated 
at the measured values of the parameters:
\begin{equation}
  f_{\rm cond}(\Omega_k \vert \{Q_i = Q_{i,\rm obs}\})
    \propto f(\Omega_k, \{Q_{i,\rm obs}\})\, ,
\end{equation}
where the constant of proportionality depends on $\{Q_{i,\rm obs}\}$, 
but not on $\Omega_k$.  Here $f_{\rm cond}$ and $f$ refer to probability 
densities for the onset of inflation.  This conditional probability 
approach does {\it not} address the question of whether these values 
$Q_{i,\rm obs}$ are in fact typical in the multiverse, i.e.\ whether 
the multiverse hypothesis is fully consistent with the current 
observations.  To study this question, we would need to estimate 
the relevant anthropic constraints on these parameters and see if 
the observed values are indeed consistent with possible underlying 
multiverse distributions.  The two approaches are complementary, 
addressing different questions.  For $\Omega_k$, we will take both 
approaches---we will study the implications of a multiverse distribution 
on future measurements assuming our current knowledge, and we will 
also ask if the predicted distribution is consistent with our current 
knowledge.

The present-day curvature contribution $\Omega_k$ is related to the 
number of $e$-folds of {\it deterministic (non-diffusive) slow-roll 
inflation} $N$.  We can thus study the probability distribution for 
$\Omega_k$ by analyzing that for $N$.  The relation between the two 
depends on the details of how the deterministic slow-roll era begins, 
but to a good approximation we can write
\begin{equation}
  \Omega_k \propto \frac{1}{e^{2N}} \, ,
\label{eq:Omega-k}
\end{equation}
provided that $\Omega_k$ is in the relevant parameter region, where 
$\Omega_k$ is smaller than $1$ but larger than the contribution 
induced by density fluctuations, $\Omega_k \simgt \delta\rho/\rho 
\approx 10^{-5}$.  The probability of observing $N$ in the interval 
between $N$ and $N+dN$ in future measurements, given our current 
knowledge, can then be written as
\begin{equation}
  P_{\rm obs}(N)\, dN 
  \propto f(N, \{Q_{i,\rm obs}\})\, M_m(N)\, C(N)\, n(N)\, dN\, .
\label{eq:P-N-2}
\end{equation}
Here $f(N, \{ Q_i \})$ is the multiversal joint probability density 
for $N$ and the set of $Q_i$, analogous to $f(\Omega_k, \{Q_i\})$. 
($f(N, \{ Q_i \})$ and $f(\Omega_k, \{Q_i\})$ are of course different 
functions with different arguments, but we use the same symbol $f$ 
because they have the same verbal description, as the joint probability 
density for a randomly chosen onset of inflation in the multiverse 
to be characterized by the arguments of $f$.)  $f(N, \{ Q_i \})$ is 
in principle determined by the statistical properties of the inflaton 
field and its potential in the multiverse.  $C(N)$ encodes our current 
knowledge about $N$, and $n(N)$ is the anthropic weighting factor. 
If any quantity $Q_i$ is subject to a non-negligible observational 
error, then we need to integrate that parameter over the observationally 
allowed range.

For $C(N)$, we know from cosmological observations that $N$ must be 
larger than a certain value $N_{\rm obs,min}$, corresponding to the 
maximum curvature allowed observationally, $\Omega_{k,{\rm max}} 
\simeq 0.01$~\cite{Komatsu:2008hk}.  Thus, we can take $C(N) \approx 
\theta(N - N_{\rm obs,min})$.  The value of $N_{\rm obs,min}$ depends 
on the history of our pocket universe, especially on the reheating 
temperature $T_R$, but is generically around $40~\mbox{--}~60$.  Any 
extra $e$-folds of inflation suppress $\Omega_k$ further as described 
by Eq.~(\ref{eq:Omega-k}), so
\begin{equation}
  \Omega_k \approx \frac{\Omega_{k,{\rm max}}}{e^{2(N-N_{\rm obs,min})}} 
    + O(10^{-5})\, .
\label{eq:Omega-k_2}
\end{equation}
The anthropic factor $n(N)$ can be chosen to be the expected number 
of observers per unit volume, summed over all time within the life 
of the pocket universe.  For fixed values of the parameters $Q_i$, we 
expect that $n(N)$ approaches some constant $n_{\infty}$ at large $N$, 
since the evolution of life will not be affected by very small spatial 
curvature.  As smaller values of $N$ are considered, at some point 
$|\Omega_k|$ will suddenly become large, growing by a factor of about 
$e^2 \approx 7.4$ each time $N$ decreases by $1$.  The probability 
for observers to evolve presumably decreases quickly as $|\Omega_k|$ 
becomes large, so we will also approximate this function by a step 
function:\ $n(N) \approx n_{\infty}\, \theta(N-N_{\rm anthropic})$. 
Since obviously $N_{\rm obs,min} > N_{\rm anthropic}$, we find
\begin{equation}
  P_{\rm obs}(N) \propto f(N,\{Q_{i,\rm obs}\})\, 
    M_m(N)\, \theta(N - N_{\rm obs,min})\, .
\label{eq:N_step}
\end{equation}

To discuss the consistency of our current measurements of $\Omega_k$ 
with the predictions of the multiverse hypothesis, we need to 
consider the predicted probability distribution $P_{{\rm obs},\raise 
0.8pt\hbox{$\scriptstyle \not$}\Omega_k}(N)$, defined as the conditional 
probability density given all of our current knowledge except for our 
measurements of $\Omega_k$.  When expressed in terms of $N$ instead 
of $\Omega_k$, this probability distribution is obtained from 
Eq.~(\ref{eq:P-N-2}) by omitting the factor $C(N)$.  Given our 
approximation for $n(N)$, we find
\begin{equation}
  P_{{\rm obs},\raise 0.8pt\hbox{$\scriptstyle \not$}\Omega_k}(N)
    \propto f(N,\{Q_{i,\rm obs}\})\, 
    M_m(N)\, \theta(N - N_{\rm anthropic})\, .
\label{eq:P-N-NoOmega}
\end{equation}
Using this probability distribution, we can check whether the probability 
of obtaining $N > N_{\rm obs,min}$ is indeed reasonable or not.

\section{Statistical Distributions for the Number of {\boldmath $e$}-folds}
\label{sec:f-N}

To use Eq.~(\ref{eq:N_step}) to estimate the probability distribution 
for future measurements of $N$, we need to know $f(N,\{Q_{i,\rm obs}\})$, 
the underlying multiversal joint probability density for the onset of 
$N$ $e$-folds of inflation with the measured values $Q_{i,\rm obs}$ of 
physical parameters.  This quantity depends crucially on the history 
of our pocket universe just before the observable inflation.  In this 
section we discuss $f(N,\{Q_{i,\rm obs}\})$ for each of the four scenarios, 
(i)~--~(iv), described in Section~\ref{subsec:pre-inflation}.

\subsection{{\boldmath $f(N,\{Q_{i,\rm obs}\})$} for scenarios (i) 
 or (ii):\ new or chaotic eternal inflation}
\label{subsec:diffusive}

Suppose that the past history of our pocket universe was either scenario 
(i) or (ii); i.e., suppose that the observable era of deterministic 
slow-roll inflation was smoothly connected to a prior era of new or chaotic 
eternal inflation.  In this case we find that $f(N,\{Q_{i,\rm obs}\})$ 
is strongly peaked at very large $N$, so that the residual curvature 
contribution in the present universe is completely negligible.

The qualitative reason for this result is very simple.  At the transition 
point between eternal and non-eternal inflation, $\varphi \equiv \varphi_*$, 
the amplitude for the scalar perturbations exiting the Hubble horizon 
is of order unity:\ $Q\bigl(k(\varphi_*)\bigr) \approx 1$.  On the 
other hand, when the current horizon scale exits the Hubble horizon 
at $\varphi \equiv \varphi_0$, the perturbation amplitude is very small:\ 
$Q\bigl(k(\varphi_0)\bigr) \approx 10^{-5}$.  Since the perturbation 
amplitude changes rather slowly with $k$, this large change in $Q$ 
implies that there must have been a large number $\varDelta N$ of 
$e$-folds of slow-roll inflation between the end of eternal inflation 
and the time when the current horizon scale exited the Hubble horizon. 
We will first show this in two simple examples, and then present 
a general argument.

We first consider an example of scenario (i), using the following 
inflaton potential:
\begin{equation}
  V = V_0 - \frac{1}{2} \mu^2 \varphi^2 + \delta V(\varphi) 
\qquad
  (V_0,\, \mu^2 > 0),
\label{eq:V-new}
\end{equation}
where $\mu^2 \simlt H_I^2 \equiv V_0/3 M_{\rm Pl}^2$ to have a flat 
potential at small $\varphi$.  We also assume, for simplicity, that 
before the current horizon scale exits the Hubble horizon we can take 
$V \approx V_0$ and $V' \approx - \mu^2 \varphi$.  $\delta V(\varphi)$ 
is assumed to be negligible during this period, although later it 
controls the ending of inflation.

With the initial conditions $\varphi \approx \dot{\varphi} \approx 0$, 
the potential of Eq.~(\ref{eq:V-new}) leads to eternal inflation for 
$0 < |\varphi| < \varphi_*$, where $\varphi_*$ is determined by the 
condition that $\varDelta \varphi_{\rm qu}$ is comparable to the 
classical motion of $\varphi$ during a Hubble time, or
\begin{equation}
  \varDelta \varphi_{\rm qu} \approx \frac{H}{2\pi} 
    \approx |\dot{\varphi}_{\rm classical}|\, H^{-1} 
    \approx \frac{|V'|}{3 H^2}\, .
\label{eq:eternal}
\end{equation}
For the potential of Eq.~(\ref{eq:V-new}), this gives
\begin{equation}
  \varphi_* \approx \frac{3}{2\pi}\, \frac{H_I^3}{\mu^2}\, .
\label{eq:phi*-new}
\end{equation}
For $|\varphi| > \varphi_*$, the evolution of $\varphi$ is described 
by the classical equation of motion, which in this approximation gives 
$\varphi(t) \propto \exp\big\{ \frac{\mu^2}{3 H_I} t \big\}$.  The 
scalar perturbation amplitude for single-field slow-roll inflation 
is given by~\cite{Liddle-Lyth}
\begin{equation}
  Q(k) \equiv \frac{2}{5} \Delta_{\cal R}(k) 
    \approx \frac{1}{\sqrt{75}\,\pi\,M_{\rm Pl}^3}\, 
      \frac{V^{3/2}}{\left| V' \right|}\, ,
\label{eq:Q-general}
\end{equation}
where $V(\varphi)$ is evaluated at the value of $\varphi$ when the 
scale $k$ exits the Hubble horizon.  For the present case, one finds 
$Q(k) \approx (3/5\pi) (H_I^3/\mu^2 \varphi)$.  Observationally, 
the perturbation amplitude at the current horizon scale $Q_0 \equiv 
Q(k=H_0) \simeq 2 \times 10^{-5}$~\cite{Komatsu:2008hk}, so
\begin{equation}
  \varphi_0 \approx \frac{3}{5\pi} \frac{H_I^3}{\mu^2 Q_0} 
    \simeq 9.5 \times 10^3 \frac{H_I^3}{\mu^2}\, .
\label{eq:phi0-new}
\end{equation}
Note that our approximation $V \approx V_0$ requires that $\frac{1}{2} 
\mu^2 \varphi_0^2 \ll V_0$, which leads to the parameter restriction 
$\mu^2 \gg H_I^4/(Q_0^2 M_{\rm Pl}^2)$.  This is consistent with the 
upper bound on $\mu^2$ provided that $H_I \ll Q_0 M_{\rm Pl} \approx 
5 \times 10^{13}~{\rm GeV}$.  The scalar spectral index $n_s$ (defined 
by $Q^2 \propto k^{n_s - 1}$) is given by~\cite{Liddle-Lyth-2}
\begin{equation}
  1 - n_s = 6 \epsilon - 2 \eta\, ,
\label{eq:n_s}
\end{equation}
where the slow-roll parameters $\epsilon$ and $\eta$ are defined by
\begin{eqnarray}
  \epsilon &\equiv& \frac{M_{\rm Pl}^2}{2} \left( \frac{V'}{V} \right)^2\, ,
\\
  \eta &\equiv& M_{\rm Pl}^2 \frac{V''}{V}\, .
\label{eq:slow-roll-parms}
\end{eqnarray}
For the current system, one finds $1 - n_s = 2\mu^2/3H_I^2$. 
Since observation gives $(1 - n_s)_{\rm obs} \simeq 0.04 \pm 
0.01$~\cite{Komatsu:2008hk}, we have
\begin{equation}
  \frac{\mu^2}{H_I^2} \simeq 0.06 \pm 0.01.
\label{eq:mu-HI-new}
\end{equation}
Thus, in this model the number of $e$-folds of slow-roll inflation before 
the exit of the current horizon scale is given by
\begin{equation}
  \varDelta N \equiv N(\varphi_*)-N(\varphi_0) 
  = \frac{3 H_I^2}{\mu^2} \ln\frac{\varphi_0}{\varphi_*} 
  \approx 500\, .
\label{eq:delN-new}
\end{equation}
Note that $\varDelta N$ is a fixed, and large, number.  This implies 
that, in the present scenario, we would not have any possibility of 
observing a residual curvature contribution in the current universe.

A similar analysis can also be performed for scenario (ii), for which 
we choose the sample potential
\begin{equation}
  V = \frac{1}{2} m^2 \varphi^2 
\qquad
  (m^2 > 0).
\label{eq:V-chaotic}
\end{equation}
The field values corresponding to Eqs.~(\ref{eq:phi*-new}) and 
(\ref{eq:phi0-new}) are now
\begin{equation}
  \varphi_* \approx \sqrt{4\pi\sqrt{6}\, \frac{M_{\rm Pl}^3}{m}}
\label{eq:phi*-chaotic}
\end{equation}
and
\begin{equation}
  \varphi_0 \approx \sqrt{10\pi\sqrt{6}\, Q_0\, \frac{M_{\rm Pl}^3}{m}}\, .
\label{eq:phi0-chaotic}
\end{equation}
The parameters are then determined uniquely by the value of $n_s$, since 
for this potential $\epsilon = \eta = 2 M_{\rm Pl}^2/\varphi^2$ and 
therefore $1 - n_{s,0} = 8 M_{\rm Pl}^2/\varphi_0^2$.  Using the observed 
values of $n_{s,0}$ and $Q_0$, one has $\varphi_0 \approx 14\, M_{\rm Pl} 
\simeq 3.4 \times 10^{19}~{\rm GeV}$, $m \approx 7.7 \times 10^{-6} 
M_{\rm Pl} \simeq 1.9 \times 10^{13}~{\rm GeV}$, and $\varphi_* \approx 
2.0 \times 10^3 M_{\rm Pl} \simeq 4.9 \times 10^{21}~{\rm GeV}$.  As in 
the previous case, the large difference between $\varphi_*$ and $\varphi_0$ 
implies that there must have been a large number $\varDelta N$ of $e$-folds 
of inflation between the end of eternal inflation and the Hubble horizon 
exit of the current horizon scale.  To find $\varDelta N$ we note that, 
in slow-roll approximation, this potential energy function gives
\begin{equation}
  \frac{d \varphi^2}{dN} = -\frac{1}{H} \frac{d \varphi^2}{dt} 
    \approx \frac{2\varphi}{H}\, \frac{V'}{3 H} 
    \approx 4 M_{\rm Pl}^2\, ,
\label{eq:dphi2-dN}
\end{equation}
so
\begin{equation}
  \varDelta N \approx \frac{1}{4 M_{\rm Pl}^2} 
    \left(\varphi_*^2 - \varphi_0^2 \right) 
  \approx 1.0 \times 10^6\, .
\label{eq:delN-chaotic}
\end{equation}
For this case, the number of $e$-folds of inflation is even much larger 
than the previous case, so again there is no possibility that curvature 
could be observed.

Having seen that $\varDelta N$ is very large for two special cases, 
we can now give a general argument that $\varDelta N$ is always very 
large for scenarios (i) and (ii).  By comparing Eqs.~(\ref{eq:eternal}) 
and (\ref{eq:Q-general}), one sees that the condition for the onset 
of eternal inflation, $\varphi \equiv \varphi_*$, is equivalent to 
$Q\bigl(k(\varphi_*)\bigr) \approx 2/5$.  If $Q(k)$ varies slowly, 
then there must be many $e$-folds of inflation between the point where 
$Q \approx 2/5$ and the point where $Q = Q_0 \simeq 2 \times 10^{-5}$. 
And $Q(k)$ does vary slowly, since $Q^2(k) \propto k^{n_s - 1}$, and 
observationally $n_s \simeq 0.96$, which is near to the scale-invariant 
value of $n_s = 1$.  To quantitatively relate a change in $Q$ to the 
number of $e$-folds over which it occurs, we recall that $k \propto 
e^{-N}$, where $N$ is the number of $e$-folds of inflation that have 
not yet occurred when the wave number $k$ exits the Hubble horizon. 
Thus $Q^2 \propto e^{(1-n_s) N}$, so
\begin{equation}
  dN \approx \frac{2}{1-n_s}\, d\ln Q\, .
\label{eq:N-Q}
\end{equation}
Thus for $\varphi \approx \varphi_0$ we have
\begin{equation}
  dN \approx 50 \, d\ln Q\, .
\label{eq:N-Q_0}
\end{equation}
This implies that even a fractional change in $Q$ of $O(1)$ around 
$\varphi = \varphi_0$ leads to a large number of $e$-folds.  Since 
$\ln Q$ changes by about $10$ as $\varphi$ varies from $\varphi_*$ 
to $\varphi_0$, the resulting $\varDelta N$ is very large.

The argument described above shows that in scenario (i) or (ii), 
$\varDelta N$ must be very large, i.e.\ the probability density 
$f(N,\{Q_{i,\rm obs}\})$ is peaked at values of $N$ much larger than 
$N(\varphi_0)$.  Note that this conclusion does {\it not} depend on 
the measure adopted.  Therefore, if our past history is either scenario 
(i) or (ii), the probability of observing curvature in future measurements 
is completely negligible.  To turn the argument around, if future 
measurements find a curvature contribution (beyond the $10^{-5}$ 
level), then we would learn that diffusive (new or chaotic type) 
eternal inflation did not occur in our ``immediate'' past.

\subsection{{\boldmath $f(N,\{Q_{i,\rm obs}\})$} for scenario (iii):\ 
quantum tunneling after eternal old inflation}
\label{subsec:bubble}

\begin{figure}
\begin{center}
  \input{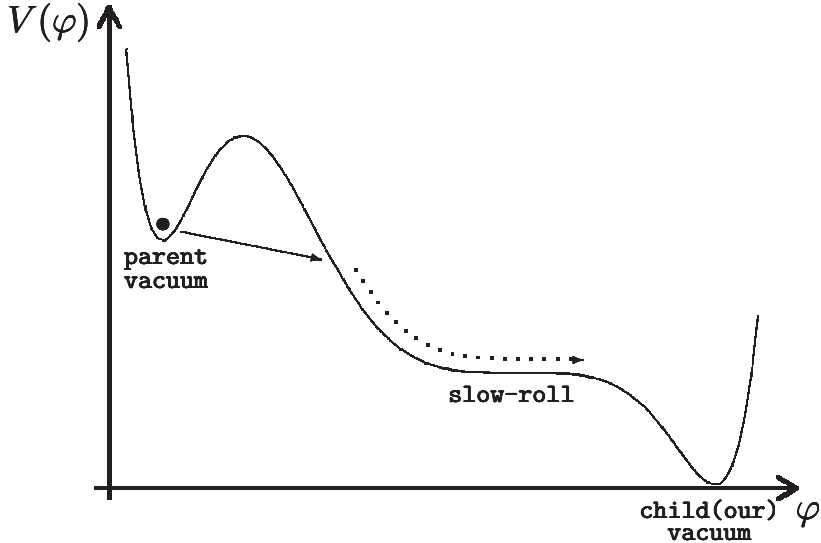}
\caption{A local neighborhood of our own vacuum in the landscape.}
\label{fig:potential}
\end{center}
\end{figure}

We now start discussion of scenario~(iii):\ eternal old inflation. 
While the previous cases could be understood solely in terms of the 
dynamics of density perturbations, for this case we will need to 
consider the description of probabilities in the multiverse.  Consider 
a diagram showing the local neighborhood of our own vacuum in the 
landscape, as depicted schematically in Fig.~\ref{fig:potential}.  The 
diagram shows a single scalar field, but it symbolically represents 
a field moving in a space with many dimensions.  We are interested 
in the situation where our pocket universe was born by a quantum 
tunneling event~\cite{Coleman:1980aw}, in which the scalar field 
$\varphi$ tunneled out from a local minimum, which we call our parent 
vacuum. The pocket universe then experienced a period of slow-roll 
inflation which ended with the scalar field rolling into the local 
minimum of our vacuum, which in this context we call a child vacuum.

We note that the transition from one vacuum to another does not always 
occur through a quantum tunneling event; if the potential barrier 
separating the two is very broad, then the field $\varphi$ climbs up 
the barrier~\cite{Hawking:1981fz}, rather than tunnels through it. 
If the transition from our parent vacuum to our vacuum occurred in 
this way, however, the field $\varphi$ started rolling into our vacuum 
from the top of the very broad barrier.  Therefore, in this case 
the situation is reduced to the one already discussed in the previous 
subsection~\cite{Batra:2006rz}.

\subsubsection{The meaning of the statistical distribution for 
{\boldmath $N$}}
\label{subsubsec:dist}

In the setup considered here, what exactly do we mean by 
$f(N,\{Q_{i,\rm obs}\})$, which we recall was described as the 
multiversal joint probability density for $N$ and $Q_i$?  In fact, if 
we focus our attention on a particular region of the landscape containing 
only one pair of parent-child vacua, as in Fig.~\ref{fig:potential}, 
then the number of $e$-folds $N$ of slow-roll inflation is just a 
fixed number, determined by the shape of the potential.  Since the 
point where fields appear after the tunneling is determined uniquely 
(at least in the semi-classical limit), there is no ``statistical 
distribution'' for $N$.%
\footnote{If the potential contains a (quasi-)flat direction around this 
 point, quantum fluctuation can give a distribution for $N$.  We ignore 
 this effect below since it is not generic.}
Nonetheless, we of course do not know the value of $N$, so we will 
describe it in terms of an estimated probability distribution, which 
includes uncertainties arising from at least two sources.

First, it is possible that the landscape includes many parent-child 
pairs that could be our pocket universe and its parent.  We would in 
fact expect that the landscape contains a large number of vacua in which 
the low-energy physical laws, including the values of the parameters, 
are consistent with what we know about our own universe.  Any one 
of these vacua would be a candidate for our local vacuum, and we 
would have no way of knowing in which one we live.  There would 
be perhaps an even larger set of vacua which tunnel to one of the 
local vacuum candidates, and we would have no way of knowing which 
of these was the parent of our pocket universe.  Since any one of 
these parent-child transitions could have been the transition that 
produced our pocket universe, the value of $N$ can acquire a statistical 
distribution.

Even if there are many parent-child pairs that could be ours, however, 
it will not lead to an actual spread in values of $N$ unless more than 
one of them occurs with significant probability.  Whether or not that 
is the case depends on branching ratios in the landscape, which is a 
topic about which little is known.  We discuss these branching ratios in 
Appendix~\ref{app:single-path}.  We do not reach a definite conclusion, 
but we find that it is not implausible that the decay rates in the 
landscape are so diverse that the decay of any given vacuum, especially 
a long-lived one, is overwhelmingly dominated by a single channel.

\begin{figure}[t]
\begin{center} 
\begin{picture}(300,220)(70,37)
  \CArc(30,185)(25,0,360)  \Text(30,185)[]{\large $C_1$}
  \Line(1.1633,131.446)(18.1475,162.988)
  \Line(18.1475,162.988)(12.4229,158.96)\Line(18.1475,162.988)(17.9355,155.991)
  \Line(25.9971,132.962)(28.0826,160.074)
  \Line(28.0826,160.074)(24.4811,154.071)\Line(28.0826,160.074)(30.7237,153.591)
  \Text(41,142)[]{\small $\cdots$}
  \Line(58.8367,131.446)(41.8525,162.988)
  \Line(41.8525,162.988)(42.0645,155.991)\Line(41.8525,162.988)(47.5771,158.96)
  \CArc(-5,120)(13,0,360) \Text(-5,120)[]{$P_{11}$}
  \CArc(25,120)(13,0,360) \Text(25,120)[]{$P_{12}$}
  \CArc(65,120)(13,0,360) \Text(65,120)[]{$P_{1n_1}$}
  \Line(-12,80)(-12,109.046)
  \Line(-12,109.046)(-15.1305,102.785)\Line(-12,109.046)(-8.8695,102.785)
  \Text(-5,93)[]{\tiny $\cdots$}
  \Line(2,80)(2,109.046)
  \Line(2,109.046)(-1.1305,102.785)\Line(2,109.046)(5.1305,102.785)
  \Line(18,80)(18,109.046)
  \Line(18,109.046)(14.8695,102.785)\Line(18,109.046)(21.1305,102.785)
  \Text(25,93)[]{\tiny $\cdots$}
  \Line(32,80)(32,109.046)
  \Line(32,109.046)(28.8695,102.785)\Line(32,109.046)(35.1305,102.785)
  \Line(58,80)(58,109.046)
  \Line(58,109.046)(54.8695,102.785)\Line(58,109.046)(61.1305,102.785)
  \Text(65,93)[]{\tiny $\cdots$}
  \Line(72,80)(72,109.046)
  \Line(72,109.046)(68.8695,102.785)\Line(72,109.046)(75.1305,102.785)
  \CArc(-15,75)(5,180,270) \Line(-15,70)(25,70) \CArc(25,65)(5,0,90)
  \CArc(75,75)(5,270,360) \Line(35,70)(75,70) \CArc(35,65)(5,90,180)
  \Text(30,59)[]{\scriptsize distribution for inflaton}
  \Text(30,51)[]{\scriptsize initial values}
  \CArc(145,185)(25,0,360) \Text(145,185)[]{\large $C_2$}
  \Line(116.1633,131.446)(133.1475,162.988)
  \Line(133.1475,162.988)(127.4229,158.96)
  \Line(133.1475,162.988)(132.9355,155.991)
  \Line(140.9971,132.962)(143.0826,160.074)
  \Line(143.0826,160.074)(139.4811,154.071)
  \Line(143.0826,160.074)(145.7237,153.591)
  \Text(156,142)[]{\small $\cdots$}
  \Line(173.8367,131.446)(156.8525,162.988)
  \Line(156.8525,162.988)(157.0645,155.991)
  \Line(156.8525,162.988)(162.5771,158.96)
  \CArc(110,120)(13,0,360) \Text(110,120)[]{$P_{21}$}
  \CArc(140,120)(13,0,360) \Text(140,120)[]{$P_{22}$}
  \CArc(180,120)(13,0,360) \Text(180,120)[]{$P_{2n_2}$}
  \Line(103,80)(103,109.046)
  \Line(103,109.046)(99.8695,102.785)\Line(103,109.046)(106.1305,102.785)
  \Text(110,93)[]{\tiny $\cdots$}
  \Line(117,80)(117,109.046)
  \Line(117,109.046)(113.8695,102.785)\Line(117,109.046)(120.1305,102.785)
  \Line(133,80)(133,109.046)
  \Line(133,109.046)(129.8695,102.785)\Line(133,109.046)(136.1305,102.785)
  \Text(140,93)[]{\tiny $\cdots$}
  \Line(147,80)(147,109.046)
  \Line(147,109.046)(143.8695,102.785)\Line(147,109.046)(150.1305,102.785)
  \Line(173,80)(173,109.046)
  \Line(173,109.046)(169.8695,102.785)\Line(173,109.046)(176.1305,102.785)
  \Text(180,93)[]{\tiny $\cdots$}
  \Line(187,80)(187,109.046)
  \Line(187,109.046)(183.8695,102.785)\Line(187,109.046)(190.1305,102.785)
  \CArc(100,75)(5,180,270) \Line(100,70)(140,70) \CArc(140,65)(5,0,90)
  \CArc(190,75)(5,270,360) \Line(150,70)(190,70) \CArc(150,65)(5,90,180)
  \Text(225,185)[]{$\cdots \cdots$}
  \CArc(305,185)(25,0,360) \Text(305,185)[]{\large $C_n$}
  \Line(276.1633,131.446)(293.1475,162.988)
  \Line(293.1475,162.988)(287.4229,158.96)
  \Line(293.1475,162.988)(292.9355,155.991)
  \Line(300.9971,132.962)(303.0826,160.074)
  \Line(303.0826,160.074)(299.4811,154.071)
  \Line(303.0826,160.074)(305.7237,153.591)
  \Text(316,142)[]{\small $\cdots$}
  \Line(333.8367,131.446)(316.8525,162.988)
  \Line(316.8525,162.988)(317.0645,155.991)
  \Line(316.8525,162.988)(322.5771,158.96)
  \CArc(270,120)(13,0,360) \Text(270,120)[]{$P_{n1}$}
  \CArc(300,120)(13,0,360) \Text(300,120)[]{$P_{n2}$}
  \CArc(340,120)(13,0,360) \Text(340,120)[]{$P_{nn_n}$}
  \Line(263,80)(263,109.046)
  \Line(263,109.046)(259.8695,102.785)\Line(263,109.046)(266.1305,102.785)
  \Text(270,93)[]{\tiny $\cdots$}
  \Line(277,80)(277,109.046)
  \Line(277,109.046)(273.8695,102.785)\Line(277,109.046)(280.1305,102.785)
  \Line(293,80)(293,109.046)
  \Line(293,109.046)(289.8695,102.785)\Line(293,109.046)(296.1305,102.785)
  \Text(300,93)[]{\tiny $\cdots$}
  \Line(307,80)(307,109.046)
  \Line(307,109.046)(303.8695,102.785)\Line(307,109.046)(310.1305,102.785)
  \Line(333,80)(333,109.046)
  \Line(333,109.046)(329.8695,102.785)\Line(333,109.046)(336.1305,102.785)
  \Text(340,93)[]{\tiny $\cdots$}
  \Line(347,80)(347,109.046)
  \Line(347,109.046)(343.8695,102.785)\Line(347,109.046)(350.1305,102.785)
  \CArc(260,75)(5,180,270) \Line(260,70)(300,70) \CArc(300,65)(5,0,90)
  \CArc(350,75)(5,270,360) \Line(310,70)(350,70) \CArc(310,65)(5,90,180)
  \CArc(0,210)(10,90,180) \Line(0,220)(157.5,220) \CArc(157.5,230)(10,270,360)
  \CArc(335,210)(10,0,90) \Line(177.5,220)(335,220) \CArc(177.5,230)(10,180,270)
  \Text(167.5,240)[]{\small distribution for inflaton potentials}
  \Text(365,185)[l]{$C_i$: {\small child (our) vacua}}
  \Text(375,120)[l]{$P_{ij}$: {\small parent vacua}}
\end{picture}
\caption{A schematic picture for a landscape leading to probability 
 distributions for inflaton potentials and initial values.}
\label{fig:model-1}
\end{center}
\end{figure}
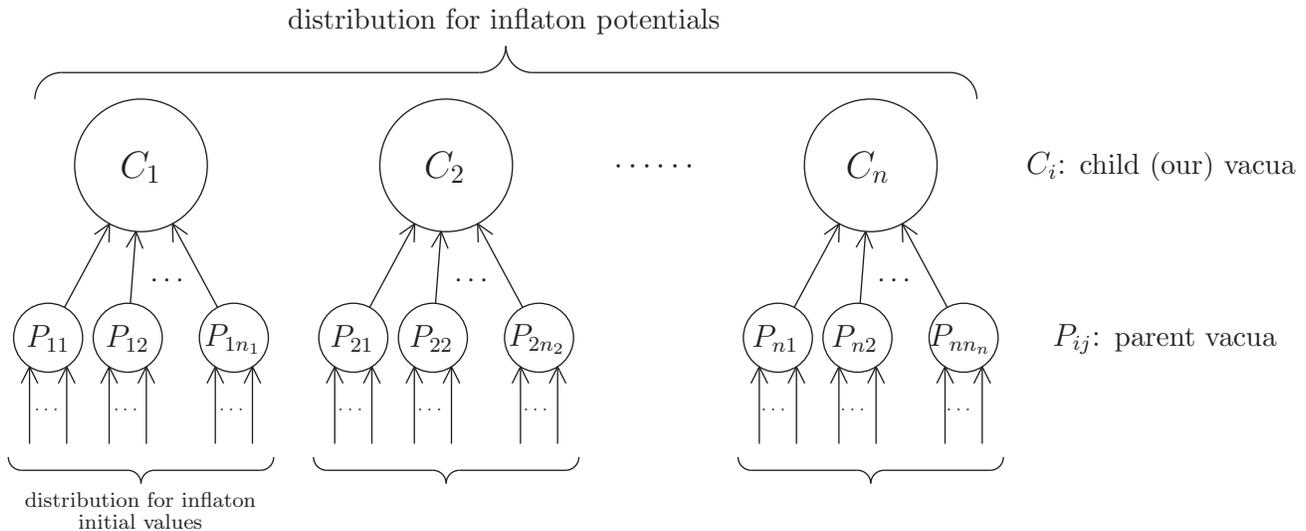
The fastest decays are most likely the least diverse, so one plausible 
scenario is that a significant fraction of the multiverse evolves 
through one or more short-lived Planck-scale vacua, which decay into 
a large number of ``second generation'' vacua with nonnegligible 
branching ratios.  Then, even if the subsequent decays are each 
dominated by a single channel, the large number of second generation 
vacua could lead to many vacua which are compatible with ours, all 
occurring with comparable probabilities.  This situation is illustrated 
in Fig.~\ref{fig:model-1}.  It leads to a probability distribution in 
$N$ because pocket universes entirely consistent with what we know about 
ours are produced by many different parent to child transitions, each 
with its own value of $N$.

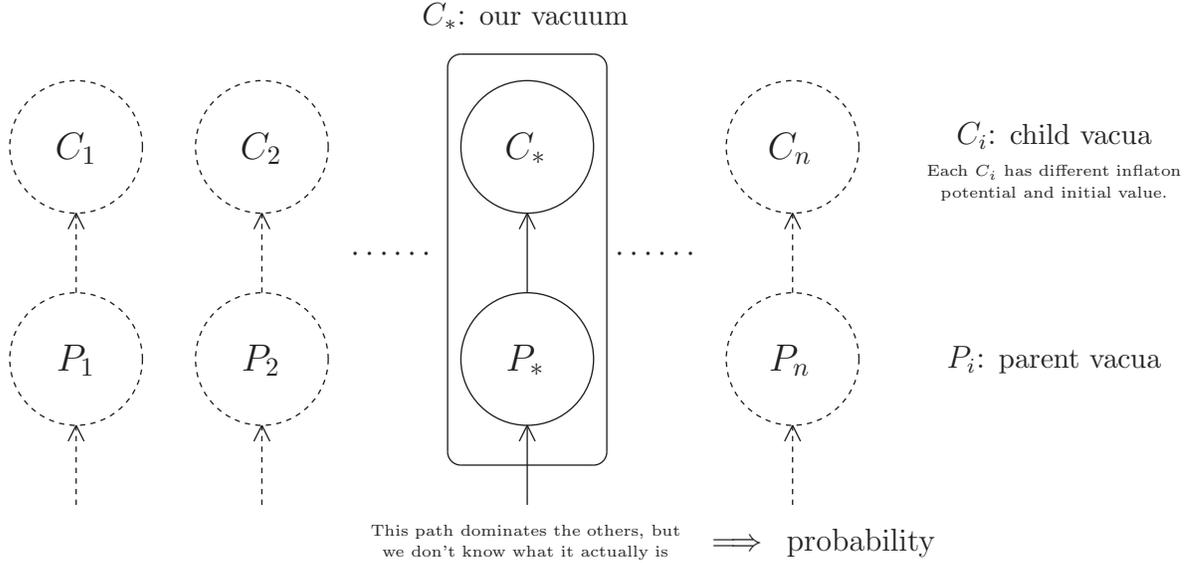
\begin{figure}[t]
\begin{center}
\begin{picture}(300,230)(30,20)
  \DashLine(0,50)(0,80){2}
  \Line(0,80)(-3,74)   \Line(0,80)(3,74)
  \DashCArc(0,105)(25,0,360){2} \Text(0,105)[]{\large $P_1$}
  \DashLine(0,130)(0,160){2}
  \Line(0,160)(-3,154)   \Line(0,160)(3,154)
  \DashCArc(0,185)(25,0,360){2} \Text(0,185)[]{\large $C_1$}
  \DashLine(70,50)(70,80){2}
  \Line(70,80)(67,74)   \Line(70,80)(73,74)
  \DashCArc(70,105)(25,0,360){2} \Text(70,105)[]{\large $P_2$}
  \DashLine(70,130)(70,160){2}
  \Line(70,160)(67,154)   \Line(70,160)(73,154)
  \DashCArc(70,185)(25,0,360){2} \Text(70,185)[]{\large $C_2$}
  \Text(120,145)[]{$\cdots\cdots$}
  \Line(170,50)(170,80)
  \Line(170,80)(167,74)   \Line(170,80)(173,74)
  \CArc(170,105)(25,0,360) \Text(170,105)[]{\large $P_*$}
  \Line(170,130)(170,160)
  \Line(170,160)(167,154)  \Line(170,160)(173,154)
  \CArc(170,185)(25,0,360) \Text(170,185)[]{\large $C_*$}
  \CArc(145,70)(5,180,270) \CArc(195,70)(5,270,360)
  \CArc(145,215)(5,90,180) \CArc(195,215)(5,0,90)
  \Line(145,65)(195,65)    \Line(145,220)(195,220)
  \Line(140,70)(140,215)   \Line(200,70)(200,215)
  \Text(170,235)[]{$C_*$: {\small our vacuum}}
  \Text(220,145)[]{$\cdots\cdots$}
  \DashLine(270,50)(270,80){2}
  \Line(270,80)(267,74)   \Line(270,80)(273,74)
  \DashCArc(270,105)(25,0,360){2} \Text(270,105)[]{\large $P_n$}
  \DashLine(270,130)(270,160){2}
  \Line(270,160)(267,154) \Line(270,160)(273,154)
  \DashCArc(270,185)(25,0,360){2} \Text(270,185)[]{\large $C_n$}
  \Text(370,190)[]{$C_i$: {\small child vacua}}
  \Text(370,176)[]{\tiny Each $C_i$ has different inflaton}
  \Text(370,168)[]{\tiny potential and initial value.}
  \Text(370,105)[]{$P_i$: {\small parent vacua}}
  \Text(170,40)[]{\tiny This path dominates the others, but}
  \Text(170,32)[]{\tiny we don't know what it actually is}
  \Text(250,35)[]{$\Longrightarrow$}
  \Text(297,36)[]{probability}
\end{picture}
\caption{A schematic picture for a landscape in which various vacuum decay 
 chains have enormously different probabilities.}
\label{fig:model-2}
\end{center}
\end{figure}
On the other hand, we can also imagine that estimates of the spread 
of decay rates in the landscape, like the ones in our preliminary 
discussion in Appendix~\ref{app:single-path}, will show that absolutely 
every decay in the landscape is almost certainly dominated by a single 
channel.  In that case, of all the vacua that are compatible with ours, 
we would expect one to completely dominate the probability.  Furthermore, 
the appearance of this vacuum would be completely dominated by the 
decay of a single type of parent vacuum.  In this situation $N$ would 
have a unique value, so the previous discussion of a probability 
distribution does not apply.  This brings us to the second source 
of uncertainty, which is ignorance.  Even if we conclude that single 
paths dominate the evolution of the multiverse, we will still not be 
capable of identifying the vacuum and parent that dominate the probability. 
We would therefore {\it parameterize our ignorance} about the most 
likely path in the form of a probability distribution for $N$.  This 
situation is illustrated in Fig.~\ref{fig:model-2}.  Note that this 
is a different concept from the probability distribution of the physical 
realization of different values of $N$ in the multiverse---in fact, 
it is closer to the concept of probability used in conventional arguments 
for naturalness in a single vacuum theory.  If we have no precise 
knowledge about the vacuum population mechanism, we are limited 
to making plausible assumptions about the probability distribution 
for $N$.

In either of the cases discussed above, the implications for future 
measurements of $\Omega_k$ are encoded in the probability distribution 
$f(N,\{Q_{i,\rm obs}\})$, as in Eq.~(\ref{eq:N_step}).  This distribution 
corresponds to the multiversal joint probability distribution for the 
onset of $N$ $e$-folds of slow-roll inflation, and the measured parameters 
$Q_i=Q_{i,\rm obs}$, introduced in section~\ref{subsec:measures}.  We 
will estimate it in the next two subsections.  Our estimate does not 
depend much on the origin of this probability, whether it represents 
physical realizations in the multiverse or the parameterization of our 
ignorance.  We therefore conclude that even if nonzero curvature is 
someday measured, this measurement will not tell us whether different 
values of $N$ are actually realized with nonnegligible probabilities 
in the multiverse.

\subsubsection{Probability distribution for the inflaton potential and 
 the starting point of slow-roll inflation}
\label{subsubsec:stat}

Our goal is to evaluate $f(N,\{Q_{i,\rm obs}\})$ for scenario (iii), 
where slow-roll inflation follows a quantum tunneling event.  We have 
in mind a potential of the form of Fig.~\ref{fig:potential}, the form 
of which leads immediately to an important issue.  The tunneling rate 
for a given transition depends on the properties of the potential 
function in the region of the barrier, while the number $N$ of $e$-folds 
of slow-roll inflation depends on the properties of the potential in 
the slow-roll part of the potential energy curve.  We do not know to 
what extent these two parts of the potential are correlated, but it is 
conceivable that the statistics of the slow-roll part of the potential 
could be strongly affected by the fact that some shapes are more likely 
to occur with a barrier that gives faster tunneling.  If the correlation 
is strong, then it is potentially a large effect, since the tunneling 
rate depends exponentially on the parameters.

Since we do not know how to calculate the correlations, we consider 
two extreme possibilities.  If these regions are only weakly 
correlated, then the tunneling rate will have no significant effect 
on $f(N,\{Q_{i,\rm obs}\})$.  If, however, the correlation is strong, 
then it could have a large effect, the nature of which we will discuss 
in the following section.

For now we write $f(N,\{Q_{i,\rm obs}\})$ as the product of two factors,
\begin{equation}
  f(N,\{Q_{i,\rm obs}\}) = f_0(N,\{Q_{i,\rm obs}\}) \,
     B(N,\{Q_{i,\rm obs}\}) \, ,
\label{eq:f_0B}
\end{equation}
where $f_0(N,\{Q_{i,\rm obs}\})$ is the answer that we would expect 
in the absence of correlations, and $B(N,\{Q_{i,\rm obs}\})$ is the 
correction factor caused by the bias toward slow-roll potentials that 
correspond to faster decay rates.  $f_0(N,\{Q_{i,\rm obs}\})$ can be 
called the vacuum statistics probability distribution, and it can be 
defined more precisely by imagining that we first make a list of all 
the parent vacua $P_\alpha$ that occur in the multiverse.  To weight 
each vacuum according to its relevance to the evolution of the multiverse, 
we imagine assigning each vacuum $P_\alpha$ a weight $W_\alpha$, which 
we take to be proportional to the relative number of nucleation events 
in which bubbles of $P_\alpha$ are produced (as determined according 
to the measure of choice).  The precise choice of this weighting will 
not affect our estimates, since we will assume that the decay properties 
of $P_\alpha$ are not correlated with the properties of its production, 
but we will see in the next section that this specification for $W_\alpha$ 
is particularly useful.  We further imagine that we can determine all 
the possible transitions by which each parent vacuum $P_\alpha$ can 
decay to each child vacuum $C_j$.  These transitions will presumably 
have a huge range of decay rates, but $f_0(N,\{Q_{i}\})$ is defined 
as the joint probability density for $N$ and $\{Q_{i}\}$ computed with 
all these transitions counted equally, weighted only by $W_\alpha$:
\begin{equation}
  f_0(N,\{Q_{i}\}) \propto \sum_{\alpha,j} \, W_\alpha \, 
    \delta\bigl(N - N(\alpha,j)\bigr) \, \prod_i \, 
    \delta\bigl(Q_i - Q_i(\alpha,j)\bigr)\, ,
\label{eq:f_0def}
\end{equation}
where $\alpha$ and $j$ are summed over all parent-child pairs, and 
$N(\alpha,j)$ and $Q_i(\alpha,j)$ are the values of the number $N$ 
of slow-roll $e$-folds and the value of measurable quantity $Q_i$ 
associated with this parent-child combination.  The constant of 
proportionality is determined by requiring $f_0(N,\{Q_{i}\})$ to 
be normalized, and $f(N,\{Q_{i,\rm obs}\})$ is obtained by setting 
each $Q_i$ to its observed value $Q_{i,{\rm obs}}$.

In this section we will estimate $f_0(N,\{Q_{i,\rm obs}\})$, leaving 
the discussion of $B(N,\{Q_{i,\rm obs}\})$ until the next section.

Following the approach of Freivogel, Kleban, Rodr\'{\i}guez Mart\'{\i}nez, 
and Susskind~\cite{Freivogel:2005vv} (hereafter called FKRS), we develop 
a toy model for the slow-roll part of the potential energy curve and 
for the value of the inflaton field at the start of the slow-roll period.%
\footnote{FKRS did not discuss the possibility of correlations between 
 the tunneling and slow-roll parts of the potential, so their $P(N)$ 
 corresponds to our $f_0(N,\{Q_{i,\rm obs}\})$.}
While FKRS used the observed value of the density perturbation amplitude 
$Q_0$ as a condition, we will use both it and the observed value of the 
scalar spectral index $n_s$.  We seek only a crude approximation---which 
is the best we can do---so we make the simplest possible assumptions. 
We assume therefore that the inflaton potential, during the era of 
slow-roll inflation, is approximated by
\begin{equation}
  V = V_0 + A \varphi + \frac{1}{2} \mu^2 \varphi^2\, .
\label{eq:V-stat}
\end{equation}
We further assume that slow-roll inflation starts at $\varphi = \Delta$ 
($>0$) and ends at $\varphi = 0$ (so we take $\partial V/\partial\varphi 
> 0$ for $0 \leq \varphi \leq \Delta$, which implies $A > 0$).  In this 
section we will pursue the hypothesis that the parameters $V_0$, $A$, 
$\mu^2$, and $\Delta$ ``scan'' in the landscape, in the sense that 
they can be assumed to vary in the multiverse according to some smooth 
probability distribution function $h_0(V_0, A, \mu^2, \Delta)$.  Here 
$h_0(V_0, A, \mu^2, \Delta)$ is defined, like $f_0(N,\{Q_{i,\rm obs}\})$, 
as a vacuum statistics probability density.  That is, it is defined by 
weighting transitions by the weight $W_\alpha$ of the parent vacuum, 
but not by the decay rate, so that correlations with the tunneling 
part of the potential play no role.  We then study the resulting 
probability distribution for $N$, the number of $e$-folds of slow-roll 
inflation.  When we consider a specific example, we will choose an 
$h_0$ that is flat.

Keeping in mind that we seek only a crude approximation, we assume that 
the parameters of the potential satisfy
\begin{equation}
  V_0 \gg A \Delta \ , \qquad V_0 \gg |\mu^2| \Delta^2 \ ,\qquad
     \hbox{and} \qquad A \gg |\mu^2 | \Delta\, .
\label{eq:inequalities}
\end{equation}
The total number of $e$-folds is then given by
\begin{equation}
  N = \int_0^\Delta \frac{V}{M_{\rm Pl}^2 V'} d\varphi 
    \approx \frac{V_0 \Delta}{A M_{\rm Pl}^2}\, .
\label{eq:N-stat}
\end{equation}
Under these approximations the density perturbation amplitude $Q_0$ 
and the spectral index $n_s$ are constant through the slow-roll period, 
given by
\begin{equation}
  Q_{0,{\rm obs}} = \frac{1}{\sqrt{75} \, \pi \, M_{\rm Pl}^3}\,
     \frac{V^{3/2}}{\left| V' \right|} \approx
     \frac{V_0^{3/2}}{\sqrt{75}\, \pi A M_{\rm Pl}^3}
\label{eq:Q_0-obs}
\end{equation}
and
\begin{equation}
  1 - n_{s,{\rm obs}} 
    = (6\epsilon - 2\eta) \approx M_{\rm Pl}^2
     \left(\frac{3A^2}{V_0^2}- \frac{2\mu^2}{V_0}\right)\, .
\label{eq:n_s-obs}
\end{equation}
The joint probability density $f_0$ for $N$, $Q_{0,{\rm obs}}$, and 
$n_{s,{\rm obs}}$ is then given by
\begin{eqnarray}
  f_0(N,Q_{0,{\rm obs}},n_{s, \rm obs})
    &=& \int\!\!\!\!\int\!\!\!\!\int\!\!\!\!\int\! 
    dV_0\, dA\, d\mu^2\, d\Delta\,\, 
  \delta\biggl(N - \frac{V_0 \Delta}{A M_{\rm Pl}^2}\biggr)
  \delta\biggl(Q_{0,{\rm obs}} 
    - \frac{V_0^{3/2}}{\sqrt{75}\,\pi\, A M_{\rm Pl}^3} \biggr)
\nonumber\\
  && {} \qquad \times
  \delta\biggl((1 - n_{s,{\rm obs}})
    - M_{\rm Pl}^2\left[\frac{3A^2}{V_0^2}-\frac{2\mu^2}{V_0}\right]\biggr) 
  \,\, h_0(V_0, A, \mu^2, \Delta)
\\
  &=& \frac{(75 \pi^2)^3 M_{\rm Pl}^2 Q_{0,{\rm obs}}^5}{N^8}
    \int\!d\Delta\, \Delta^7\,\, 
    h_0 \biggl( \frac{75\pi^2 Q_{0,{\rm obs}}^2 \Delta^2 M_{\rm Pl}^2}{N^2}, 
      \frac{75\pi^2 Q_{0,{\rm obs}}^2 \Delta^3}{N^3},
\nonumber\\
  && {} \qquad
    \frac{75\pi^2 Q_{0,{\rm obs}}^2 \Delta^2}{2N^2} 
    \left[ \frac{3 \Delta^2}{N^2 M_{\rm Pl}^2} - (1-n_{s,{\rm obs}}) 
      \right], \Delta \biggr)\, .
\label{eq:DN-1}
\end{eqnarray}

As a simple example, we assume that the distribution $h_0(V_0, A, 
\mu^2, \Delta)$ is constant in the range $0 < V_0 < V_{0,\rmax}$, 
$0 < A < A_\rmax$, $\mu^2_\rmin < \mu^2 < \mu^2_\rmax$, and $0 < \Delta 
< \Delta_\rmax$, where $\mu^2_\rmin < 0$.  The approximations described 
by Eq.~(\ref{eq:inequalities}) are not really valid throughout this 
range, but in the spirit of our crude approximation we will ignore this 
problem.  Then the integral in Eq.~(\ref{eq:DN-1}) depends on $N$ only 
through the limit of integration:\ that is, if $N$ is sufficiently small, 
then one of the first three arguments of $h_0$ can reach its upper limit 
before $\Delta$ reaches $\Delta_\rmax$.  In this case the upper limit 
of integration becomes proportional to $N$, resulting in a factor of 
$N^8$, canceling the prefactor.  Thus
\begin{equation}
  f_0(N,Q_{0,{\rm obs}},n_{s,{\rm obs}}) 
    \propto \begin{cases} \displaystyle 
    \frac{1}{N^8}  &  \text{if $N > N_\rmin\,$,}\\
    {\rm const.}   &  \text{if $N < N_\rmin\,$,} 
  \end{cases}
\label{eq:f(N)cases}
\end{equation}
where
\setbox0=\hbox{$\displaystyle \left\{ \Biggl( 
    \frac{A^2_\rmax}{\Bigl[ \sqrt{\bigl[ \bigr]^2} \Bigr]} 
  \Biggr)^{1/2} \right\} $}  
\begin{eqnarray}
  N_\rmin &=& \max \left\{ 
    {\vrule height \ht0 depth \dp0 width 0pt} 
    \frac{\sqrt{75}\,\pi M_{\rm Pl} Q_{0,{\rm obs}}
      \Delta_\rmax}{\sqrt{V_{0,\rmax}}}\, , \
    \left( \frac{75\pi^2 Q_{0,{\rm obs}}^2}{A_\rmax} \right)^{1/3} 
    \Delta_\rmax\, ,\right.
\nonumber\\
  && \hskip -20pt \left. \Biggl( \frac{6\sqrt{75}\, \pi Q_{0,{\rm obs}} 
      \Delta_\rmax^2}{M_{\rm Pl} \Bigl[ \sqrt{ \bigl[\sqrt{75}\,\pi 
      M_{\rm Pl} Q_{0,{\rm obs}} (1-n_s) \bigr]^2 + 24 \mu^2_\rmax} 
    + \sqrt{75}\,\pi M_{\rm Pl} Q_{0,{\rm obs}} (1-n_s) \Bigr]} \Biggr)^{1/2} 
      \right\}\, .
\label{eq:Nmin}
\end{eqnarray}
The arguments of the $\max$ function in the above expression are the 
values of $N$ for which each of the first three arguments of $h_0$, 
in Eq.~(\ref{eq:DN-1}), will reach its maximum value before $\Delta$ 
reaches $\Delta_\rmax$.%
\footnote{To be complete, there is one further complication that could 
 occur, but which we assume does not occur.  For small $\Delta$, the 
 $\mu^2$ argument (i.e, the 3rd argument) of $h$ in Eq.~(\ref{eq:DN-1}) 
 can be negative, so the integration can be limited by $\mu^2_\rmin$, 
 the smallest allowed value of $\mu^2$.  We will assume, however, that 
 $\mu^2_\rmin$ is chosen to be sufficiently negative to prevent this 
 from happening.  The minimum possible value for this argument is 
 $-75 \pi^2 M_{\rm Pl}^2 (1-n_{s,{\rm obs}})^2 Q_{0,{\rm obs}}^2/24$, 
 which is small because $Q_{0,{\rm obs}} \simeq 2 \times 10^{-5}$, 
 so one can easily choose $\mu^2_\rmin$ to avoid this complication.}

Equation~(\ref{eq:Nmin}) is very complicated, but fortunately all we 
really need to know is that $N_\rmin$ is generically small.  For sample 
values we can take all the integration limits to be at the Planck scale:\ 
i.e., $V_{0,\rmax} = M_{\rm Pl}^4$, $A_\rmax = M_{\rm Pl}^3$, $\mu^2_\rmax 
= -\mu^2_\rmin = M_{\rm Pl}^2$, and $\Delta_\rmax = M_{\rm Pl}$.  Then 
with the measured values of $Q_{0,{\rm obs}}$ and $n_{s,{\rm obs}}$, 
the three arguments in Eq.~(\ref{eq:Nmin}) become $0.00054$, $0.0067$, 
and $0.026$, respectively.  Thus, for all interesting values of $N$, 
this example gives $f_0(N,Q_{0,{\rm obs}},n_{s,{\rm obs}}) \propto 1/N^8$.

There are many variants of this analysis, however, so we do not claim 
that there is any particular significance to the power 8.  If we had 
not conditioned on $n_{s,{\rm obs}}$, whether or not we included the 
$\mu^2$ term in the potential, we would have found $f_0(N,Q_{0,{\rm obs}}) 
\propto 1/N^6$.  (In this case $N_\rmin$ would be larger than before, 
based on the first two arguments of $h_0$, but it would still be less 
than 1 for the Planck-scale sample values.)  We might also consider 
omitting the $\mu^2$ term from the potential, but conditioning on 
$n_{s,{\rm obs}}$ nonetheless.  In that case the power counting gives 
a probability density that is flat, but one also finds that the arguments 
of $h_0$ become crucial.  The value of $\Delta$ will be forced outside 
the allowed range unless $N < N_\rmax$, where
\begin{equation}
  N_\rmax = \frac{\sqrt{3} \Delta_\rmax}{M_{\rm Pl} 
     \sqrt{1-n_{s,{\rm obs}}}} \, .
\label{eq:Nmax}
\end{equation}
For the Planck-scale sample values this gives $N_\rmax = 8.7$, although 
it can be moved up to the interesting range if we allow $\Delta_\rmax$ 
to be a few times larger than $M_{\rm Pl}$.

FKRS used a different parameterization of the potential,
\begin{equation}
  V(\varphi) = V_0(1-x \varphi/\Delta) \, . 
\label{eq:FKRSpotential}
\end{equation}
Assuming a flat probability distribution for $V_0$, $x$, and $\Delta$, 
and by conditioning on $Q_{0,{\rm obs}}$ but not $n_{s,{\rm obs}}$, 
they found that $f_0(N,Q_{0,{\rm obs}}) \propto 1/N^4$.  They did not 
specify a range of validity for this result, but we find that it is 
valid for $N > N_\rmin$, where
\begin{equation}
  N_\rmin = \max \left( \frac{\sqrt{75}\,\pi\, M_{\rm Pl} Q_{0,{\rm obs}} 
    \Delta_\rmax}{\sqrt{V_{0, \rmax}}}\ ,
  \frac{\Delta_\rmax^2}{M_{\rm Pl}^2 \, x_\rmax} \right) \, .
\label{eq:Nmin-2}
\end{equation}
For Planck-scale sample values, with $x_\rmax = 1$ as used by FKRS, 
this gives $N_\rmin = 1$, coming from the second argument of the 
$\max$ function.  While these estimates give $N_\rmin \ll 40$, for 
the FKRS parameterization it is not unreasonable to consider values 
of $\Delta_\rmax$ and $x_\rmax$ for which $N_\rmin$ might be larger 
than 60.  In that case $f_0(N,Q_{0,{\rm obs}})$ would fall as 
$1/N^{3/2}$ in the range of interest.

If one conditions on both $Q_{0,{\rm obs}}$ and $n_{s,{\rm obs}}$, 
using Eq.~(\ref{eq:FKRSpotential}) and a flat probability density for 
$V_0$, $x$, and $\Delta$, one finds $f_0(N,Q_{0,{\rm obs}},n_{s,{\rm obs}}) 
\propto N$ for $N < N_\rmax$, but $f_0(N,Q_{0,{\rm obs}},n_{s,{\rm obs}}) 
= 0$ for $N > N_\rmax$, where
\begin{equation}
  N_\rmax = \min \left( \frac{3 x_\rmax}{1- n_{s,{\rm obs}}} \ ,
    \frac{\sqrt{3}\, \Delta_\rmax}{M_{\rm Pl} 
    \sqrt{1 -n_{s,{\rm obs}}}} \right)\, .
\end{equation}
As with Eq.~(\ref{eq:Nmax}), for Planck-scale sample values this gives 
$N_\rmax = 8.7$, from the second argument of the $\min$ function. 
Again it can be increased if we allow $\Delta_\rmax$ to be larger 
than $M_{\rm Pl}$.

One can also consider adding a $\frac{1}{2} \mu^2 \varphi^2$ term 
to the potential of Eq.~(\ref{eq:FKRSpotential}), assigning a flat 
probability density to $\mu^2$ along with the other parameters. 
If one does not condition on $n_{s,{\rm obs}}$, then with our 
approximations the addition of the $\mu^2$ term has no effect on 
$f_0(N,Q_{0,{\rm obs}})$.  If one does condition on $n_{s,{\rm obs}}$, 
then $f_0(N,Q_{0,{\rm obs}},n_{s,{\rm obs}}) \propto 1/N^6$, provided 
that $N > N_\rmin$, where $N_\rmin$ is the max of both arguments in 
Eq.~(\ref{eq:Nmin-2}) and the last argument of Eq.~(\ref{eq:Nmin}).

The details of these results are of course not to be trusted, since 
they are based on ad hoc assumptions about the probability distribution 
for potential functions in the multiverse.  Nonetheless, we believe that 
we can reasonably infer that the function $f_0(N,\{Q_{i,\rm obs}\})$, 
as defined by Eq.~(\ref{eq:f_0def}), can be taken as
\begin{equation}
  f_0(N,\{Q_{i,{\rm obs}}\}) \propto \frac{1}{N^p}
\label{eq:DN-3}
\end{equation}
for some (small) power $p > 0$.  Here, the positivity of $p$ represents 
the improbability of finding an inflaton potential that supports many 
$e$-folds of evolution with a value of $Q_0$ as small as $2 \times 10^{-5}$. 
This result is mostly in agreement with FKRS, who find $f(N,Q_{0,{\rm obs}}) 
\propto 1/N^4$, except that we allow for the possibility that there 
\emph{might} be a significant correction factor $B(N, \{Q_{i,\rm obs}\})$, 
as in Eq.~(\ref{eq:f_0B}), caused by correlations with the tunneling rate.

Equation~(\ref{eq:DN-3}) is the generic behavior, but there is a plausible 
exception.  Suppose there is a mechanism which ensures the flatness of 
the inflaton potential in the vicinity of our (child) vacua:\ for example, 
a (softly broken) shift symmetry acting on the inflaton field $\varphi$. 
In terms of the model potential of Eq.~(\ref{eq:V-stat}), such a mechanism 
would ensure that $A$ is very small.  By combining Eqs.~(\ref{eq:N-stat}) 
and (\ref{eq:Q_0-obs}), the number of $e$-folds of inflation can be 
written as
\begin{equation}
  N(A,\Delta) = \left( \frac{75 \pi^2 Q_{0,{\rm obs}}^2}{A} 
    \right)^{1/3} \Delta\, ,
\label{eq:N(A,Delta)}
\end{equation}
which shows how large values of $N$ result from small values of $A$. 
In most situations the probability of finding large values of $N$ is 
suppressed by the need to find unusually small values of $A$, but a 
mechanism such as a shift symmetry can avoid that problem.  If the 
mechanism makes it probable to find values of $A$ so small that 
$N(A,\Delta) \simgt 60$ for $\Delta < \Delta_\rmax$, then we would 
expect the suppression of large $N$ would be removed.  The results 
we obtained in Eqs.~(\ref{eq:f(N)cases}) and (\ref{eq:Nmin}) verify 
these expectations, if we describe the mechanism as one that enforces 
a very small value of $A_\rmax$.  By comparing Eq.~(\ref{eq:N(A,Delta)}) 
with the second argument of Eq.~(\ref{eq:Nmin}), we see that if $A_\rmax$ 
is small enough to allow 60 $e$-folds of inflation, then $N_\rmin \ge 
60$, and then Eq.~(\ref{eq:f(N)cases}) implies that we are on the flat 
part of the probability density curve.  Thus, a mechanism to ensure 
the flatness of the potential can lead to
\begin{equation}
  f_0(N,\{Q_{i,\rm obs}\}) \sim {\rm const.}
\label{eq:DN-4}
\end{equation}
for the relevant range of $N$, so the preference to shorter inflation 
in Eq.~(\ref{eq:DN-3}) does not arise.  In fact, the consideration 
here can be used to discriminate if the observable inflation arose 
``accidentally,'' which leads to Eq.~(\ref{eq:DN-3}), or due to some 
mechanism:\ if nonzero curvature is measured, this would be strong 
evidence against a mechanism that ensures a flat potential.

Finally, although it is not needed for the main arguments presented in 
this paper, it is interesting to use the probability distributions that 
have been modeled in this section to ask what is the absolute probability 
of finding instances of inflation like the one that apparently began 
our pocket universe.  Specifically, we can use the models discussed 
in this section to calculate the probability $\bar{P}_1$ that a given 
instance of inflation will satisfy $N > \bar{N}$, $Q_0 < \bar{Q}_0$, 
and $|1 - n_s| < \varDelta \bar{n}_s$.  Here we set the bias correction 
factor $B(N,\{Q_{i,\rm obs}\}) = 1$; in the following section we will 
see that $B$ can decrease $\bar{P}_1$, but for most choices of measure 
it cannot increase it.  For the model used in Eq.~(\ref{eq:DN-1}), we 
can assume that $\bar{N} > N_\rmin$, and then the integration extends 
to $\Delta = \Delta_\rmax$, giving a factor $\Delta_\rmax^8/8$.  Using 
a normalized flat probability density for $h_0(V_0,A,\mu^2,\Delta)$, 
the probability described above is given by
\begin{eqnarray}
  \bar{P}_1
  &=& \int_{\bar{N}}^\infty\! dN\, \int_0^{\bar{Q}_0}\! dQ_0\, 
    \int_{1 - \varDelta\bar{n}_s}^{1 + \varDelta \bar{n}_s}\! dn_s\, 
    f_0(N, Q_0, n_s)
\nonumber\\
  &=& \frac{(75 \pi^2)^3 M_{\rm Pl}^2 \Delta_\rmax^8}
    {8 V_{0,{\rm max}} A_\rmax \Delta_\rmax (\mu^2_\rmax - \mu^2_\rmin)} 
    \int_{\bar{N}}^\infty\! dN\, \int_0^{\bar{Q}_0}\! dQ_0\, 
    \int_{1 - \varDelta \bar{n}_s}^{1 + \varDelta \bar{n}_s}\! dn_s\, 
    \frac{Q_0^5}{N^8}
\nonumber\\
  &=& \frac{(75 \pi^2)^3 M_{\rm Pl}^2 \Delta_\rmax^7 \bar{Q}_0^6 
    \varDelta\bar{n}_s}{168 V_{0,{\rm max}} A_\rmax (\mu^2_\rmax - 
    \mu^2_\rmin) \bar{N}^7}\, .
\end{eqnarray}
If we take $\bar{N} = 60$, $\bar{Q}_0 = Q_{0,{\rm obs}} = 2 \times 
10^{-5}$, $\varDelta\bar{n}_s = 0.04$, and Planck-scale parameters 
for the probability distribution, we find $\bar{P}_1 = 1.1 \times 
10^{-36}$.  If instead we ask for the probability that $N > \bar{N}$ 
and $Q_0 < \bar{Q}_0$, without specifying $n_s$, then we find
\begin{equation}
  \bar{P}_2 = \frac{(75 \pi^2)^2 M_{\rm Pl}^2 \Delta_\rmax^5
    \bar{Q}_0^4}{60 V_{0,{\rm max}} A_\rmax \bar{N}^5}\, ,
\end{equation}
which is valid whether or not the $\frac{1}{2} \mu^2 \varphi^2$ term 
is included in the potential.  For the parameters specified above, 
this gives $\bar{P}_2 = 1.9 \times 10^{-24}$.

For comparison, the same questions can be answered using the FKRS 
parameterization, and the associated flat probability distribution 
in $V_0$, $x$, $\Delta$, and possibly $\mu^2$.  If we include the 
$\frac{1}{2} \mu^2 \varphi^2$ term and ask for the probability 
$\bar{P}_1'$ that a given instance will satisfy $N > \bar{N}$, 
$Q_0 < \bar{Q}_0$, and $|1 - n_s| < \varDelta\bar{n}_s$, we find
\begin{equation}
  \bar{P}_1' = \frac{(75 \pi^2)^2 \Delta_\rmax^6 \bar{Q}_0^4 
    \varDelta\bar{n}_s}{70 V_{0,{\rm max}} x_\rmax (\mu^2_\rmax 
    - \mu^2_\rmin) \bar{N}^5}\, .
\end{equation}
For the numbers used above, this evaluates to $\bar{P}_1' = 3.2 \times 
10^{-26}$.  If instead we exclude the $\frac{1}{2} \mu^2 \varphi^2$ 
term and ask for the probability $\bar{P}_2'$ that $N > \bar{N}$ and 
$Q_0 < \bar{Q}_0$, without specifying $n_s$, then we find
\begin{equation}
  \bar{P}_2' = \frac{75\pi^2 \bar{Q}_0^2 \Delta_\rmax^4}{15 
    V_{0,{\rm max}} x_\rmax \bar{N}^3}\, ,
\end{equation}
which, for the numbers used here, is equal to $\bar{P}_2' = 9.1 \times 
10^{-14}$.

The detailed answers here depend very much on the ad hoc assumptions, 
and are not to be trusted, but the thrust of the answers is clear. 
First, in this picture the probability of seeing an episode of inflation 
that is suitable to begin our pocket universe is very small.  The key 
point is that 60 $e$-folds is large compared to one $e$-fold, and 
$Q_{0,{\rm obs}} \simeq 2 \times 10^{-5}$ is small compared to one. 
But we have assumed probability distributions that in no way favor 
large numbers of $e$-folds or small $Q_{0,{\rm obs}}$, so the required 
values are found only in a small corner of the probability space.  This 
feature could be changed dramatically if the underlying theory incorporated 
some mechanism to favor the right kind of potential, as we discussed 
at Eq.~(\ref{eq:DN-4}).  Nonetheless, it is certainly not clear that 
any such probability enhancement is needed for the picture to be viable, 
because with $10^{500}$ or more vacua estimated to exist in the landscape, 
probabilities like $10^{-36}$ are very large.  We would expect the 
landscape to contain a colossal number of possibilities for inflation 
to occur in exactly the right way to produce our pocket universe.  One 
then argues that there are selection effects that explain why we would 
expect to find ourselves living in such a pocket universe.  FKRS argue 
that at least 59.5 $e$-folds of inflation are necessary to explain the 
evolution of structure even at only the level of dwarf galaxies, and 
that with this condition the probability of having at least 62 $e$-folds, 
which is enough to explain the observed homogeneity and flatness, 
is high:\ about 88\%.  We will examine this question in 
Section~\ref{sec:expectation}, finding similar results.

\subsubsection{The role of nucleation rates in the statistical 
 distribution of {\boldmath $N$}}
\label{subsubsec:exp}

In Eq.~(\ref{eq:f_0B}) we expressed $f(N,\{Q_{i,\rm obs}\})$---the joint 
probability density for the number $N$ of $e$-folds of slow-roll inflation 
and the measured parameters $\{Q_{i}\}$ for a bubble universe consistent 
with our observations and arising from a randomly selected quantum 
tunneling event---as the product of two factors, $f_0(N,\{Q_{i,\rm obs}\})$ 
and $B(N,\{Q_{i,\rm obs}\})$.  $f_0(N,\{Q_{i,\rm obs}\})$ is the vacuum 
statistics probability density, given by Eq.~(\ref{eq:f_0def}), defined 
so that correlations between $N$ and the tunneling rate are irrelevant. 
$B(N,\{Q_{i,\rm obs}\})$ is the factor that corrects for any bias caused 
by the correlations with the tunneling rate, and it is the purpose of 
this section to estimate this factor.

As described at the beginning of the previous section, we are not aware 
of any way of estimating the strength of the correlations between the 
tunneling region of the potential and the slow-roll region, so we are 
allowing for the two extreme possibilities.  If these correlations are 
very weak, then $B=1$.  The rest of this section will be concerned with 
estimating $B$ when the correlations are strong.

A transition can be described by specifying the parent $P_\alpha$ and 
the child $C_j$.  For each such transition there is a nucleation rate 
$\lambda_{j\alpha}$, the number of tunneling events per physical spacetime 
volume.  To understand the bias factor $B$, we need to understand the 
relation between $\lambda_{j\alpha}$ and the probability $p_{j\alpha}$ 
that a randomly chosen quantum tunneling event is of the type $P_\alpha 
\to C_j$.

Whether $p_{j\alpha}$ indeed depends on the nucleation rate 
$\lambda_{j\alpha}$ is a measure-dependent question.  Here we argue 
that for the measures in classes (II) and (III), which are the ones 
that we consider most plausible, the probability $p_{j\alpha}$ is 
unchanged by any overall change in the decay rates from $P_\alpha$, 
but is proportional to the branching ratio of the decay of $P_\alpha$ 
to $C_j$.  We first explain this result, and then discuss its 
consequences.

We begin with the measure of Ref.~\cite{Garriga:2005av}, an example 
of measures in class~(II).  This measure adopts the method of comoving 
horizon cutoff, where the probabilities are defined by the ratios 
of the number of bubbles whose comoving sizes are greater than some 
small number $\epsilon$ ($\rightarrow 0$).  The relative probability 
$p_{j\alpha}$ is then
\begin{equation}
  p_{j\alpha} \propto H_\alpha^q \kappa_{j\alpha} s_\alpha\, ,
\label{eq:CHC-1}
\end{equation}
where $H_\alpha$ is the Hubble parameter in the parent vacuum $P_\alpha$, 
$\kappa_{j\alpha} \equiv (4\pi/3) \lambda_{j\alpha} H_\alpha^{-4}$ is 
the dimensionless nucleation rate, and $q$ and $s_\alpha$ are given by 
the asymptotic behavior of the fraction of comoving volume occupied by 
a (non-terminal) vacuum $X$ at time $t$:
\begin{equation}
  f_X(t) \rightarrow s_X e^{-q t}\, .
\label{eq:CHC-2}
\end{equation}
In the above equations, we have adopted the expressions that apply 
when we take $t$ to be the scale-factor time, although the final result 
does not depend on the choice of the time variable.  The asymptotic 
behavior of Eq.~(\ref{eq:CHC-2}) is obtained by solving the rate equation
\begin{equation}
  \frac{d f_X}{dt} = \sum_Y M_{XY} f_Y\, ,
\qquad
  M_{XY} = \kappa_{XY} - \delta_{XY} \sum_Z \kappa_{ZX}\, ,
\label{eq:CHC-3}
\end{equation}
where $\kappa_{XY} = (4\pi/3) \lambda_{XY} H_Y^{-4}$, and $X$, $Y$ 
and $Z$ run over all the vacua in the landscape.  All nonzero eigenvalues 
of $M_{XY}$ have negative real parts, and the eigenvalue with the smallest 
(by magnitude) real part is pure real, and is denoted by $-q$.  This 
eigenvalue controls the asymptotic behavior of $f_X$ and appears in 
Eq.~(\ref{eq:CHC-2}).  The vector $s_X$ is proportional to the eigenvector 
of $M_{XY}$ corresponding to the eigenvalue $-q$, and is determined by
\begin{equation}
  \Bigl( \sum_Y \kappa_{YX} - q \Bigr) s_X = \sum_Z \kappa_{XZ} s_Z\, .
\label{eq:CHC-4}
\end{equation}

Equation~(\ref{eq:CHC-1}) shows formally that $p_{j\alpha} \propto 
\kappa_{j\alpha}$, but we need to be careful, because $s_\alpha$ is itself 
determined by the nucleation rates.  We will use Eq.~(\ref{eq:CHC-4}) 
to understand the dependence of $s_\alpha$ on the $\kappa_{j\alpha}$. 
We first note that the positivity of $s_X$ implies that $q$ is smaller 
than the decay rate of the slowest decaying vacuum, called the dominant 
vacuum $D$:\ $q \leq {\rm min}_X (\sum_Y \kappa_{YX}) \equiv \kappa_D$. 
In fact, assuming that upward transitions have very small rates, 
$q \approx \kappa_D$ to a very good approximation (and $s_X \approx 
\delta_{XD}$ at the leading order)~\cite{SchwartzPerlov:2006hi}. 
(Ref.~\cite{DeSimone:2008if} points out that the dominant vacuum 
could in fact be replaced by a closely spaced system of vacuum states, 
but that does not affect the conclusions here.)  Since bubble nucleation 
rates are exponentially sensitive to the parameters of the potential, 
we expect that the $-q$ term in Eq.~(\ref{eq:CHC-4}) is negligible 
except for $X = D$:
\begin{equation}
  \sum_Y \kappa_{YX} s_X = \sum_Z \kappa_{XZ} s_Z
\quad
  \mbox{ for $X \neq D$}\, .
\label{eq:CHC-5}
\end{equation}
Note that we can regard $\kappa_{YX} s_X$ as a ``probability current'' 
associated with the transition $X \rightarrow Y$, and then this equation 
is simply a statement of current conservation, where $X = D$ acts as 
a source and terminal vacua $T$, defined by $\sum_Y \kappa_{YT} = 0$, 
as sinks.

To determine the dependence of $s_\alpha$ on the $\kappa_{i\alpha}$, 
we can rewrite Eq.~(\ref{eq:CHC-5}) with a relabeling of the indices:
\begin{equation}
  \kappa_\alpha s_\alpha = \sum_Z \kappa_{\alpha Z} s_Z \, ,
\label{eq:Sumi}
\end{equation}
where $\kappa_\alpha = \sum_j \kappa_{j\alpha}$.  In both situations 
discussed in Section~\ref{subsubsec:dist}, Figs.~\ref{fig:model-1} and 
\ref{fig:model-2}, it is reasonable to expect that the history leading 
to various parent vacua $\alpha$ is statistically independent with that 
afterwards, e.g.\ how fast those vacua decay:\ $\kappa_\alpha$.  Under 
this assumption, the right-hand side of Eq.~(\ref{eq:Sumi}) can be 
taken to be independent of $\kappa_\alpha$ (at least in the sense that 
there is no statistical correlation between the right-hand side of 
Eq.~(\ref{eq:Sumi}) and $\kappa_\alpha$), leading to $s_\alpha \propto 
\kappa_\alpha^{-1}$.  Inserting this result into Eq.~(\ref{eq:CHC-1}), 
we see that
\begin{equation}
  p_{j\alpha} \propto \frac{\kappa_{j\alpha}}{\kappa_\alpha} \, ,
\label{eq:branch}
\end{equation}
which says simply that the probability of observing a transition from 
parent $P_\alpha$ to child $C_j$ is proportional to the branching ratio 
for this transition, but is unaffected by the absolute decay rate of the 
parent $P_\alpha$.  (Note that the right-hand side of Eq.~(\ref{eq:Sumi}) 
is proportional to $W_\alpha$ as defined above Eq.~(\ref{eq:f_0def}), so 
we have found that $p_{j\alpha} \propto W_\alpha \times \hbox{branching 
ratio}$, motivating the weighting used in Eq.~(\ref{eq:f_0def}).)

The dependence of $p_{j\alpha}$ on branching ratios, but not on absolute 
decay rates, can also be shown for other measures.  For the scale-factor 
cutoff measure, an example of measures in class~(III), a calculation of 
$p_{j\alpha}$ has been performed in Ref.~\cite{DeSimone:2008if}, giving 
$p_{j\alpha} \simprop \kappa_{j\alpha} s_\alpha$ where $s_\alpha$ again 
satisfies Eq.~(\ref{eq:CHC-4}).  Equation~(\ref{eq:branch}) follows 
immediately by the same argument.  Scale-factor measure can also be 
analyzed by recasting it as a local ``fat geodesic'' measure, as described 
in Ref.~\cite{Bousso:2008hz}, and then the relative numbers of different 
transitions are clearly determined only by branching ratios that are 
encountered as the fat geodesic is followed into the future.  The same 
result can be seen for the causal patch measure, also in class~(III), 
using a local formulation analogous to the fat geodesic formulation. 
Specifically, Bousso~\cite{Bousso:2006ev} has shown that the probabilities 
$p_{j\alpha}$ in the causal patch measure can be computed by following 
a single geodesic, so they are determined directly from the branching 
ratios.  The quantum measure of Ref.~\cite{Nomura:2011dt} also gives 
Eq.~(\ref{eq:branch}) (see Appendix~\ref{app:quantum}).

Thus, for a wide class of measures, the probabilities $p_{j\alpha}$ 
depend on the branching ratios of decays, but not on the decay rates 
themselves, which depend exponentially on the parameters of the potential 
energy function.  What does this tell us about the dependence of 
$f(N,\{Q_{i,\rm obs}\})$ on $N$?

As we said at the start of this section, if the correlation between the 
tunneling region and the slow-role region of the potential is very weak,
then $B=1$.  But if the correlation is strong, there are two
possibilities:\ faster tunneling rates can correlate with smaller values
of $N$, or larger values.

Consider first the case in which faster tunneling rates correlate with 
smaller values of $N$, thereby exerting pressure toward smaller $N$.  We 
do not know how to estimate the strength of the correlation, but we can 
bound the effect by considering the most extreme possibility.  Suppose, 
therefore, that for any given parent $P_\alpha$ we identify the nearest 
neighbors in the landscape, and assume that the decay rates to any other 
states are negligible.  We let $K$ be the number of such neighbors, and 
for simplicity we will assume that $K$ is the same for all vacua, with a 
value of perhaps several hundred.  For the strongest possible correlations, 
we can assume that the fastest decay will correspond to the smallest 
value of $N$, the second fastest decay will correspond to the second 
smallest value of $N$, etc., through the list of all $K$ decay modes. 
To maximize the magnitude of the effect, we will further assume that 
the decays are dominated by the fastest, so that all other decay rates 
are negligible.  (In Appendix~\ref{app:single-path} we find that this 
situation is actually quite plausible.)  Since the fastest decay is 
also the one with smallest $N$, we find that, for any parent $P_\alpha$, 
the branching ratio is $1$ for the decay with the smallest value of 
$N$, and all other branching ratios can be approximated as zero.

If we now look at the transitions $P_\alpha \to C_j$ that contribute to 
$f_0(N,\{Q_{i,\rm obs}\})$, as described by Eq.~(\ref{eq:f_0def}), we 
see that the final distribution $f(N,\{Q_{i,\rm obs}\})$ is obtained by 
examining each pair $(\alpha,j)$ and applying a test:\ if the transition 
gives the smallest value of $N$ of all $K$ decays of $P_\alpha$, then 
its branching ratio is $1$, and it is kept.  If, however, the value of 
$N$ for the transition is not the smallest of all decays of $P_\alpha$, 
the branching ratio is zero, and it is dropped.  Thus, we can obtain 
an equation for $f(N,\{Q_{i,\rm obs}\})$ by multiplying each term in 
Eq.~(\ref{eq:f_0def}) by the probability that the term corresponds to 
the lowest value of $N$ out of $K$ choices, and then renormalizing. 
But the new factor is just the probability that the other $K-1$ values 
of $N$ are larger, so for this case
\begin{equation}
  B(N,\{Q_{i,\rm obs}\}) = B_1(N,\{Q_{i,\rm obs}\}) \propto \left[ 
     \int_N^\infty \, d N \, f_0(N,\{Q_{i,\rm obs}\}) \right]^{K-1} \, .
\label{eq:B(N)-1}
\end{equation}
If $f_0(N,\{Q_{i,\rm obs}\}) \propto 1/N^p$, then $B(N,\{Q_{i,\rm 
obs}\}) \propto 1/N^{(K-1)(p-1)}$, which is a huge suppression for 
large $N$.  This was of course calculated as the maximum possible effect. 
Since we do not know how to assess the degree of correlation between 
tunneling rates and $N$, we could imagine suppression by any power 
of $N$ from zero up to $(K-1)(p-1)$.

Now consider the alternative extreme, in which faster tunneling rates
correlate with larger values of $N$, thereby exerting pressure toward
larger $N$.  The logic is all the same, but the result is very
different.  Eq.~(\ref{eq:B(N)-1}) is replaced by
\begin{equation}
  B(N,\{Q_{i,\rm obs}\}) = B_2(N,\{Q_{i,\rm obs}\}) \propto \left[
     \int_0^N \, d N \, f_0(N,\{Q_{i,\rm obs}\}) \right]^{K-1} \, .
\label{eq:B(N)-2}
\end{equation}
Only the limits integration are different, but because of the fact that
$f_0(N,\{Q_{i,\rm obs}\})$ strongly favors small $N$, the quantity in
square brackets now has a value very close to $1$ for interesting values
of $N$.  Raising it to a large power does not produce a big effect. 
In fact, if we use Eq.~(\ref{eq:f(N)cases}) as an example, and choose
$N=60$ and $K=300$, we find $B(N,\{Q_{i,\rm obs}\})=1$ to better than 20
decimal places!  For those measures for which only the branching ratios
are relevant, correlations between tunneling rates and $N$ cannot drive
$N$ to larger values.  The reason is simply that if $N$ is near 60, it
is almost certainly the largest $N$ among all the decays of the parent,
so requiring it to be the largest has no effect.

While it is plausible that a given vacuum can have significant decay 
rates to only a few hundred nearest neighbors, we would like to also 
allow for the possibility that this is wrong, and that maybe a significant 
fraction of the landscape is available as a potentially significant 
decay channel.  In that case we should take $K$ to be 10 to the power 
of several hundred, and the whole picture changes.  Then the powers 
in Eqs.~(\ref{eq:B(N)-1}) and (\ref{eq:B(N)-2}) become enormous, and 
the factors in square brackets become completely controlling.  As we 
will see in Appendix~\ref{app:single-path}, if we choose the largest 
or smallest element out of $\sim 10^{500}$ tries, from a normal 
distribution, the result is expected to be 48 standard deviations 
away from the mean.  If there is a perfect correlation between $N$ 
and the tunneling rate, as we assumed in the extreme example above, 
then $N$ would be driven effectively to 0 or infinity, and the 
situation would have already been ruled out (if $N$ is driven to 0), 
or else $N$ would be essentially infinite.  More realistically, however, 
we only know that choosing the fastest decay out of something like 
$10^{500}$ possibilities will result in a decay rate with an action 
that is of order 50 standard deviations smaller than the mean, but 
the strength of the correlation with $N$ is unknown.  The probability 
distribution for $N$ could, therefore, be biased in either direction, 
and the bias might be weak or strong.

While measures of classes (II) and (III) generically lead to probabilities 
that depend only on branching ratios, as in Eq.~(\ref{eq:branch}), not 
quite all measures of interest fit this description.  In particular, 
the stationary measure of Ref.~\cite{Linde:2007nm} does not really fit 
into our classifications; it has many properties of class (I), while at 
the same time avoiding the youngness paradox.  For measures in classes 
(II) and (III) we have seen that the factor of $\kappa_{j\alpha}$ 
appearing in Eq.~(\ref{eq:CHC-1}) is accompanied by $s_\alpha \propto 
\kappa_\alpha^{-1}$, but that happens only when the abundances of the 
potential parent vacua $P_\alpha$ are determined primarily by their 
decay rates.  In measures of class (I) it is the production rate of 
a given vacuum that primarily determines its abundance, while the decay 
rate of $P_\alpha$ has almost no effect on its abundance.  Then there 
is no factor of $\kappa_\alpha^{-1}$ accompanying $\kappa_{j\alpha}$, 
and the probability of observing a transition from any parent $P_\alpha$ 
to any child $C_i$ is proportional to $\kappa_{j\alpha}$.  Since decay 
rates behave exponentially in the parameters of the potential function, 
for the stationary measure we expect that $B(N,\{Q_{i,\rm obs}\})$ 
can be written as
\begin{equation}
  B(N,\{Q_{i,\rm obs}\}) = e^{\beta(N,\{Q_{i,\rm obs}\})}\, ,
\label{eq:fN-exp}
\end{equation}
where $\beta(N,\{Q_{i,\rm obs}\})$ is a mild, non-exponential function of 
$N$.  Thus, $f(N,\{Q_{i,\rm obs}\})$ can have an exponential sensitivity 
to $N$.  As in the class (II) and (III) cases, however, we could 
conjecture that the tunneling rates are only very weakly correlated 
with the slow-roll part of the potential, in which case we have $B=1$, 
or equivalently $\beta=0$, as before.  In any case, the final probability 
$P(N)$ for stationary measure certainly has the slow-roll volume 
increase factor $e^{3N}$, i.e.\ Eq.~(\ref{eq:P-N_class-2}).  We 
are not aware of any measure in which $f(N)$ takes the form of 
Eq.~(\ref{eq:fN-exp}) while $P(N)$ does not depend on the slow-roll 
volume increase factor, $e^{3N}$.

\subsubsection{{\boldmath $f(N,\{Q_{i,\rm obs}\})$} for scenario (iii), 
 quantum tunneling:\ summary of results}
\label{subsubsec:dist-N}

We now summarize what we have learned about the probability distribution 
of $N$ in a bubble universe.  According to the discussions in the previous 
two subsections, the probability density $f(N,\{Q_{i,\rm obs}\})$ for 
the {\it onset} of an episode of slow-roll inflation that is compatible 
with our observations $\{Q_{i,\rm obs}\}$ can be written generically as
\begin{equation}
  f(N,\{Q_{i,\rm obs}\}) \propto \frac{1}{N^p}\,e^{\beta(N)} \, . 
  \label{eq:fN-comb}
\end{equation}
If the slow-roll part of the potential is correlated with the part 
controlling the tunneling, and if the correlation favors small values 
of $N$, then we might generically expect
\begin{equation}
  p \gg 1\ .
\label{eq:p-corr}
\end{equation}
The effects of correlations between the slow-roll and tunneling 
parts of the potential for some measures depend crucially on how many 
significant decay channels compete in the decay of a given vacuum. 
Perhaps only a few hundred nearest neighbors in the landscape are 
relevant (small $K$ option), or perhaps a substantial fraction of 
the landscape is relevant (large $K$ option).  If there is only weak 
correlation between the slow-roll part of the potential and the part 
controlling the tunneling, or if the correlation favors large values 
of $N$ and we are considering the small $K$ option, then the power 
$p$ is determined purely by the statistics for the slow-roll part of 
the potential; then the analysis of Section~\ref{subsubsec:stat} gives
\begin{equation}
  p\, \left\{ 
    \begin{array}{l}
      \! > 0 \\
      \! = 0 
  \end{array} \right.
  \mbox{if the observable inflation occurs}
\left\{ 
    \begin{array}{l}
      \mbox{\!\!accidentally} \\
      \mbox{\!\!due to some mechanism\ .}
  \end{array} \right.
\label{eq:p}
\end{equation}
With the large $K$ option, correlations in the potential that favor large 
$N$ can be very significant; they need not have a power-law behavior.

The exponent $\beta(N)$ has the possibility of being nonzero only for 
measures, such as the stationary measure, which have the property that 
the probability of observing a transition from $P_\alpha$ to $C_j$ 
depends on the tunneling rate for the transition, and not just the 
branching ratio.  For measures of this type, $\beta(N)$ can arise 
due to correlations between the slow-roll part of the potential and 
the part controlling the tunneling; if those correlations are weak, 
then $\beta(N) \approx 0$.  But if the correlations are strong, then 
$\beta(N)$ can be very significant.  Since we know very little about 
$\beta(N)$, we can use the fact that we are interested in only a small 
range of $N$ about $N_{\rm obs,min}$ to expand $\beta(N)$ in a Taylor 
series:\ $\beta(N) \approx \beta(N_{\rm obs,min}) + \{ \partial\beta/\partial 
N(N_{\rm obs,min}) \} (N-N_{\rm obs,min}) + O(N-N_{\rm obs,min})^2$. 
The constant term does not affect the dependence on $N$, so we can 
drop it and replace $\beta(N)$ in Eq.~(\ref{eq:fN-comb}) by
\begin{equation}
  \beta(N) \rightarrow \beta' N\, ,
\label{eq:beta-N}
\end{equation}
where $\beta' \equiv \partial\beta/\partial N(N_{\rm obs,min})$ is 
a constant that does not depend on $N$, which can take either sign 
depending on details of the landscape potential.  The magnitude of 
$\beta'$ can be as small as zero, but to estimate how large it might 
be, we recall that it arises from the correlation between $N$ and 
the tunneling rate $\Gamma \sim e^{-S}$ of the parent-to-child vacuum 
decay, where $S$ is the bounce action associated with the decay. 
Suppose the potential barrier separating the parent from child vacuum 
is characterized by a field distance $\varDelta \varphi$, a barrier 
height $\varDelta V_h$, and an energy density difference $\varDelta 
V_{\rm diff}$.  Then, the bounce action generically scales as the thin 
wall limit expression $S \approx \frac{27 \pi^2}{2} \varDelta \varphi^4\, 
\varDelta V_h^2 / \varDelta V_{\rm diff}^3$~\cite{Coleman:1977py}. 
If the amount of slow-roll inflation $N$ is strongly correlated with 
the part of the potential energy function relevant for the tunneling, 
then we might estimate $|\beta'| \sim O(S/N_{\rm obs,min})$:
\begin{equation}
  |\beta'| \sim 
    O\left( \frac{27 \pi^2}{2} \frac{\varDelta \varphi^4 \, \varDelta 
    V_h^2}{N_{\rm obs,min} \, \varDelta V_{\rm diff}^3} \right)\, ,
\label{eq:beta'}
\end{equation}
which can easily be much larger than 1, depending on parameters.  In 
our estimation $|\beta'|$ can lie anywhere from zero up to a number 
of the order shown above, and it can have either sign.

The probability density $P_{\rm obs}(N)$ of finding ourselves in 
a region that has undergone $N$ $e$-folds of slow-roll inflation 
is then given by Eq.~(\ref{eq:N_step}), where $M_m(N)$ (defined by 
Eq.~(\ref{eq:P-N-M})) depends on whether the measure rewards volume 
increase by slow-roll inflation (class (II)) or not (class (III)). 
We therefore obtain
\begin{equation}
  P_{\rm obs}(N) \propto \frac{1}{N^p}\, e^{q N}\, 
    \theta(N-N_{\rm obs,min})\, ,
\label{eq:PN-final}
\end{equation}
where $p$ is given by Eq.~(\ref{eq:p-corr}) or (\ref{eq:p}), and $q 
= 3 + \beta'$ for class~(II) measures while $q = \beta'$ for class~(III) 
measures; these values for $p$ and $q$ are summarized in Table~\ref{tab:q}. 
This is the expression we will use in our phenomenological analysis in 
Section~\ref{sec:expectation}.
\begin{table}[t]
\begin{center}
Class (II) measures: rewarding slow-roll volume increase
\vspace{2mm}\\
\begin{tabular}{|c|c|c|}
  \hline
    & 
    \begin{tabular}{c} Measures not depending \\ 
         on decay rates \end{tabular} & 
    \begin{tabular}{c} Measures depending \\ 
         on decay rates \end{tabular} \\ 
  \hline
    \begin{tabular}{c} Weak correlation \\ 
         between slow-roll and tunneling \end{tabular} & 
    $p \geq 0$;\, $q = 3$ & 
    $p \geq 0$;\, $q \simeq 3$ \\ 
  \hline
    \begin{tabular}{c} Strong positive correlation \\ 
         between slow-roll and tunneling \end{tabular} & 
    $p \geq 0$; $q = 3$ & 
    \begin{tabular}{c} $p \geq 0$;\, $q = 3 + \beta'$\\ 
         \noalign{\vskip 2pt}
         $\beta' \simgt O(1)$\end{tabular} \\ 
  \hline
    \begin{tabular}{c} Strong negative correlation \\ 
         between slow-roll and tunneling \end{tabular} & 
     $p \gg 1$; $q = 3$ & 
     \begin{tabular}{c} $p \gg 1$;\, $q = 3 + \beta'$\\ 
         \noalign{\vskip 2pt}
         $-\beta' \simgt O(1)$\end{tabular} \\
\hline
\end{tabular}
\end{center}
\begin{center}
Class (III) measures: not rewarding volume increase
\vspace{2mm}\\
\begin{tabular}{|c|c|c|}
  \hline
    & 
    \begin{tabular}{c} Measures not depending \\ 
         on decay rates \end{tabular} & 
    \begin{tabular}{c} Measures depending \\ 
         on decay rates \end{tabular} \\ 
  \hline
    \begin{tabular}{c} Weak correlation \\ 
         between slow-roll and tunneling \end{tabular} & 
    $p \geq 0$;\, $q = 0$ & 
    $p \geq 0$;\, $|q| \ll 1$ \\ 
  \hline
    \begin{tabular}{c} Strong positive correlation \\ 
         between slow-roll and tunneling \end{tabular} & 
    $p \geq 0$;\, $q = 0$ & 
    \begin{tabular}{c} $p \geq 0$;\, $q = \beta'$\\ 
         \noalign{\vskip 2pt}
         $\beta' \simgt O(1)$\end{tabular} \\
  \hline
    \begin{tabular}{c} Strong negative correlation \\ 
         between slow-roll and tunneling \end{tabular} & 
    $p \gg 1$;\, $q = 0$ & 
    \begin{tabular}{c} $p \gg 1$;\, $q = \beta'$\\ 
         \noalign{\vskip 2pt}
         $-\beta' \simgt O(1)$\end{tabular} \\
\hline
\end{tabular}
\end{center}
\caption{Expected values of $p$ and $q$ in Eq.~(\ref{eq:PN-final}) for 
 class (II) and (III) measures.  They depend on whether the slow-roll 
 and tunneling parts of the potential are weakly or strongly correlated, 
 and on whether the correlation is positive (favoring large values of 
 $N$) or negative (favoring small values of $N$).  They also depend 
 on whether the measure predicts that the probability of observing 
 a particular transition depends only on its branching ratio (middle 
 column), or depends on the decay rate (right column).  The table is 
 constructed for the small $K$ option (see the text).  The large $K$ 
 option would change the behavior of strong positive correlations for 
 measures not depending on decay rates, giving a strong push toward 
 large $N$ which is not necessarily a power law.}
\label{tab:q}
\end{table}

\subsection{{\boldmath $f(N,\{Q_{i,\rm obs}\})$} for scenario (iv):\ 
 inflation preceded by a prior episode of inflation}
\label{subsec:sharp}

We finally consider scenario (iv), the case where there is another 
episode of inflation just before our last cosmic inflation.  In this 
case, the power spectrum of density fluctuations, ${\cal P}(k)$, can 
show a sharp spike as a function of the momentum scale $k$.  One might, 
therefore, think that this can provide a nonzero curvature over the 
visible universe, either positive or negative, by having large fluctuations 
at a length scale beyond the current horizon.  This is, however, not 
the case.  Since low multipoles of CMB temperature fluctuations are 
sensitive to density fluctuations at scales larger than the horizon 
(Grishchuk-Zel'dovich effect~\cite{Grishchuk:1978}), the observed size 
of these low multipoles $\approx 10^{-5}$ does not allow the curvature 
to extend much beyond $|\Omega_k| \approx O(10^{-5})$~\cite{Kleban}.

Therefore, even if the past history of our pocket universe is 
complicated so that ${\cal P}(k)$ has a nontrivial structure, we 
do not expect to see curvature coming from density fluctuations 
at a level, e.g., beyond $|\Omega_k| \approx 10^{-4}$.%
\footnote{While completing this paper, Ref.~\cite{Kleban:2012ph} has 
 appeared which quantitatively analyzes this issue, finding that the 
 probability of obtaining $|\Omega_k| > 10^{-4}$ from superhorizon 
 density fluctuations in a model consistent with the CMB is less than 
 $\approx 10^{-6}$.}
Since we know of no other way that positive curvature can be generated 
in a multiverse model, we conclude that a future measurement of positive 
curvature at a level of $\Omega_k \simlt -10^{-4}$ would exclude 
the entire framework considered here.  Any observation of negative 
curvature at $\Omega_k \simgt 10^{-4}$ would have to be attributed 
to Coleman-De~Luccia tunneling.

\section{Expectations for the Number of {\boldmath $e$}-folds and Curvature}
\label{sec:expectation}

\begin{figure}[t]
 \center{
 \subfigure[Consistency of current observations]
   {\includegraphics{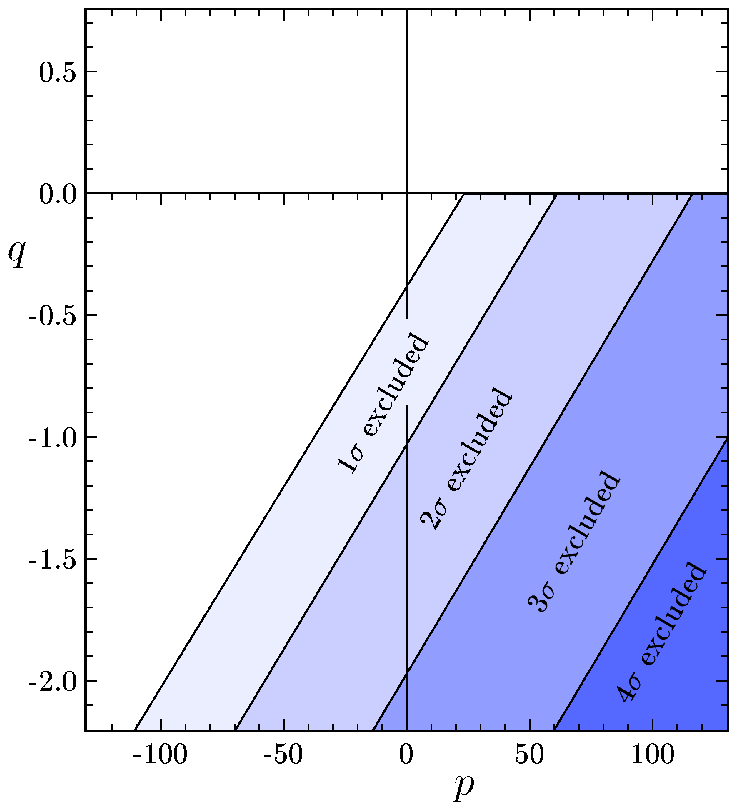}}
 \subfigure[Probabilities for future observations]
   {\includegraphics{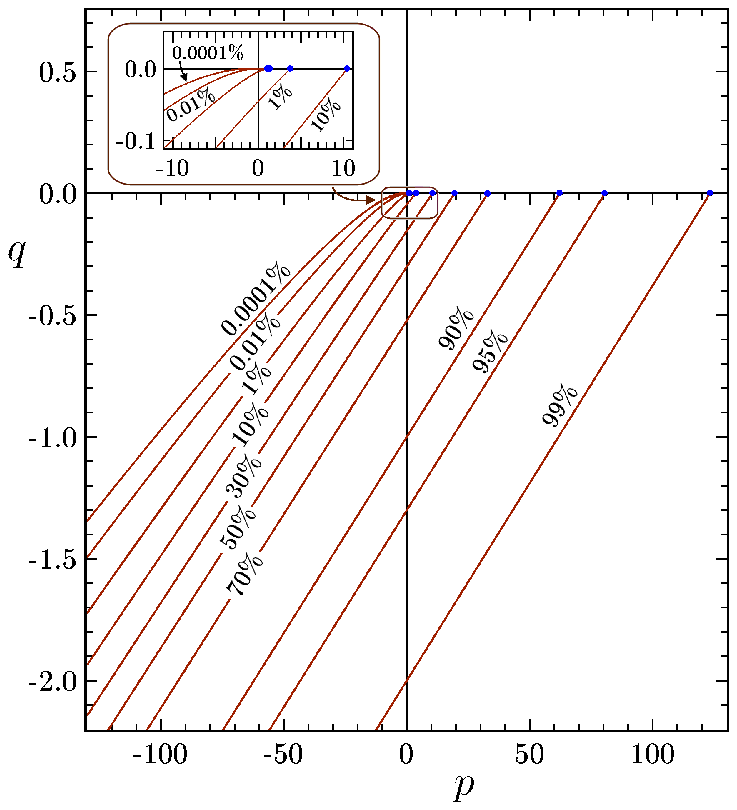}}}
 \caption{The diagram on the left shows the consistency of the current 
 bound $\Omega_k \simlt 0.01$ with multiverse probabilities.  Assuming 
 a probability distribution for the number $N$ of $e$-folds of slow-roll 
 inflation given by $P(N) \propto e^{q N}/N^p$, we calculate the probability 
 that a pocket universe which has undergone $N_{\rm anthropic}=57$ 
 $e$-folds of inflation will experience at least $N_{\rm obs,min}=60$ 
 $e$-folds.  The hypothesis that our pocket universe was drawn from 
 such a probability distribution would be excluded at the $1\sigma$, 
 $2\sigma$, $3\sigma$, or $4\sigma$ level if this probability is less 
 than $31.7\%$, $4.6\%$, $0.27\%$, or $0.0063\%$, respectively.  The 
 diagram shows the excluded regions in the $p$-$q$ plane.  Under the 
 same assumptions, the diagram on the right shows the probability that 
 our pocket universe has $\Omega_k > 10^{-4}$.  More precisely, it 
 shows the probability that a pocket universe which has undergone 60 
 $e$-folds of slow-roll inflation will not inflate by more than another 
 factor of 10 (thereby suppressing $\Omega_k$ by no more than another 
 factor of 100).}
\label{fig:p-q-contours}
\end{figure}

We now discuss implications of the probability distribution in 
Eq.~(\ref{eq:PN-final}) for current and future measurements of 
curvature.  Recall that $N_{\rm obs,min}$ denotes the minimum amount 
of slow-roll inflation required to satisfy the current observational 
constraint, $\Omega_k \simlt 0.01$.  (In this section we consider only 
$\Omega_k > 0$.)  Its value depends on the detailed history of our own 
pocket universe, especially on the reheating temperature, but is in 
the range $N_{\rm obs,min} \approx (40~\mbox{--}~60)$.  Following 
FKRS, we assume that the requirement of structure formation provides 
an anthropic lower bound on the amount of slow-roll inflation:
\begin{equation}
  N_{\rm anthropic} \simeq N_{\rm obs,min} - 3.0\, .
\label{eq:N_anthropic}
\end{equation}
(Here, we have assumed only the weak requirement that dwarf galaxies form. 
If we require that typical galaxies form, then $3.0$ is replaced by $1.9$.) 
To test the consistency of the current constraint on $\Omega_k$ with 
multiverse probabilities, we use Eqs.~(\ref{eq:P-N-NoOmega}) with 
$f(N,\{Q_{i,\rm obs}\}) M_m(N) = e^{q N}/N^p$ (see Eq.~(\ref{eq:PN-final})) 
to express the probability $P_{\rm current}$ that a pocket universe which 
has undergone $N_{\rm anthropic}$ $e$-folds of slow-roll inflation will 
go on to undergo at least $N_{\rm obs,min}$ $e$-folds of inflation:
\begin{eqnarray}
  P_{\rm current} &=& \int_{N_{\rm obs,min}}^\infty \! dN \,
     P_{{\rm obs},\raise 0.8pt\hbox{$\scriptstyle \not$}\Omega_k}(N)
\nonumber\\
  &=& \left. \int_{N_{\rm obs,min}}^\infty \! dN \, 
      \frac{e^{q N}}{N^p} \right/ 
      \int_{N_{\rm anthropic}}^\infty \! dN \, 
      \frac{e^{q N}}{N^p} \, .
\label{eq:P_current}
\end{eqnarray}
Figure~\ref{fig:p-q-contours}(a) shows which regions of the $p$-$q$ 
plane are excluded by yielding low values of $P_{\rm current}$, at 
various levels of confidence, using $N_{\rm obs,min} = 60$.

Figure~\ref{fig:p-q-contours}(a) shows that $q > 0$ is always allowed, 
but we should keep in mind that this is based on an idealization that is 
not reliable.  It arises from the fact that the probability distribution 
$P(N) \propto e^{q N}/N^p$ diverges at large $N$ for any $q>0$, for any 
value of $p$.  But if $q$ is positive and small, and $p$ is positive 
and large, then the divergent behavior will not occur until $N$ is very 
large, at which point the linear approximation that we introduced in 
Eq.~(\ref{eq:beta-N}) will no longer be valid.  Thus, for small positive 
$q$ and large positive $p$, a more sophisticated analysis would be needed.

If we assume that there is only a weak correlation between the tunneling 
and slow-roll parts of the potential function, then measures of class 
(II), which reward slow-roll volume increases, are clearly allowed by 
Fig.~\ref{fig:p-q-contours}(a).  As shown in Table~\ref{tab:q}, these 
measures give $q = 3$ or at least $q \simeq 3$.  Measures of class (III), 
which do not reward slow-roll volume increases, are also consistent with 
$P_{\rm current}$.  These measures give $q$ either equal to zero or very 
small, so the graph shows that the hypothesis is excluded at the $1\sigma$ 
level only if $p \simgt 23$.  By contrast, in Section~\ref{subsubsec:stat} 
we found that values in the range of $p=0$ to $p=8$ seemed plausible.

If there is a strong, positive correlation (i.e., favoring large $N$) 
between the tunneling and slow-roll parts of the potential function, 
then all the measures shown in Table~\ref{tab:q} are again consistent 
with $P_{\rm current}$.  For measures that do not depend on decay rates, 
for the small $K$ option (as defined in Section~\ref{subsubsec:dist-N}), 
the situation is identical to that described in the previous paragraph; 
for the large $K$ option, the pressure toward large $N$ improves the 
consistency.  For those measures that depend on decay rates, $q$ is 
given a positive contribution $\beta'$ of order $1$, which pushes an 
already acceptable $(p,q)$ combination further from the excluded regions.

If, however, there is a strong negative correlation (i.e., favoring 
small $N$) between the tunneling and slow-roll parts of the potential 
function, then measures of class (III) (not rewarding volume increases) 
are very likely excluded, depending on exactly how strong the correlations 
are.  The correlations cause $p$ to become large, and for measures 
depending on decay rates, $q$ to become negative as well.  Only the 
mildest range of ``strong'' negative correlations would be consistent. 
Measures of class (II), which reward slow-roll volume increase, 
would still be allowed if they do not depend on decay rates, since 
they would have $q=3$.  But for those that do depend on decay rates, 
$q=3+\beta'$, where $\beta'<0$, so it could be allowed or not, depending 
on the magnitude of $\beta'$.

To discuss future measurements, we note that our pocket universe will 
have a curvature beyond $\Omega_k$ if the amount of slow-roll inflation 
satisfies
\begin{equation}
  N < N_{\rm obs,min} + \frac{1}{2} 
    \ln\frac{\Omega_{k,{\rm max}}}{\Omega_k} 
  \equiv N(\Omega_k)\, ,
\label{eq:N-obs}
\end{equation}
where $\Omega_{k,{\rm max}} \simeq 0.01$ is the maximum curvature allowed 
by the current observation.  Recalling Eq.~(\ref{eq:PN-final}) for the 
probability density for the number $N$ of $e$-folds of slow-roll inflation 
experienced by our pocket universe, the probability that $N < N(\Omega_k)$ 
is given by
\begin{eqnarray}
  P_{\rm future}(\Omega_k) &=& \int_0^{N(\Omega_k)}\! dN\, 
    P_{\rm obs}(N)
\nonumber\\
  &=& \left. \int_{N_{\rm obs,min}}^{N(\Omega_k)} \! dN\, 
     \frac{e^{q N}}{N^p} \right/ 
     \int_{N_{\rm obs,min}}^\infty \! dN\, 
     \frac{e^{q N}}{N^p} \, .
\label{eq:P_future}
\end{eqnarray}
\begin{figure}[t]
  \subfigure[$N_{\rm obs,min} = 60$]{\includegraphics[width=8.5cm]{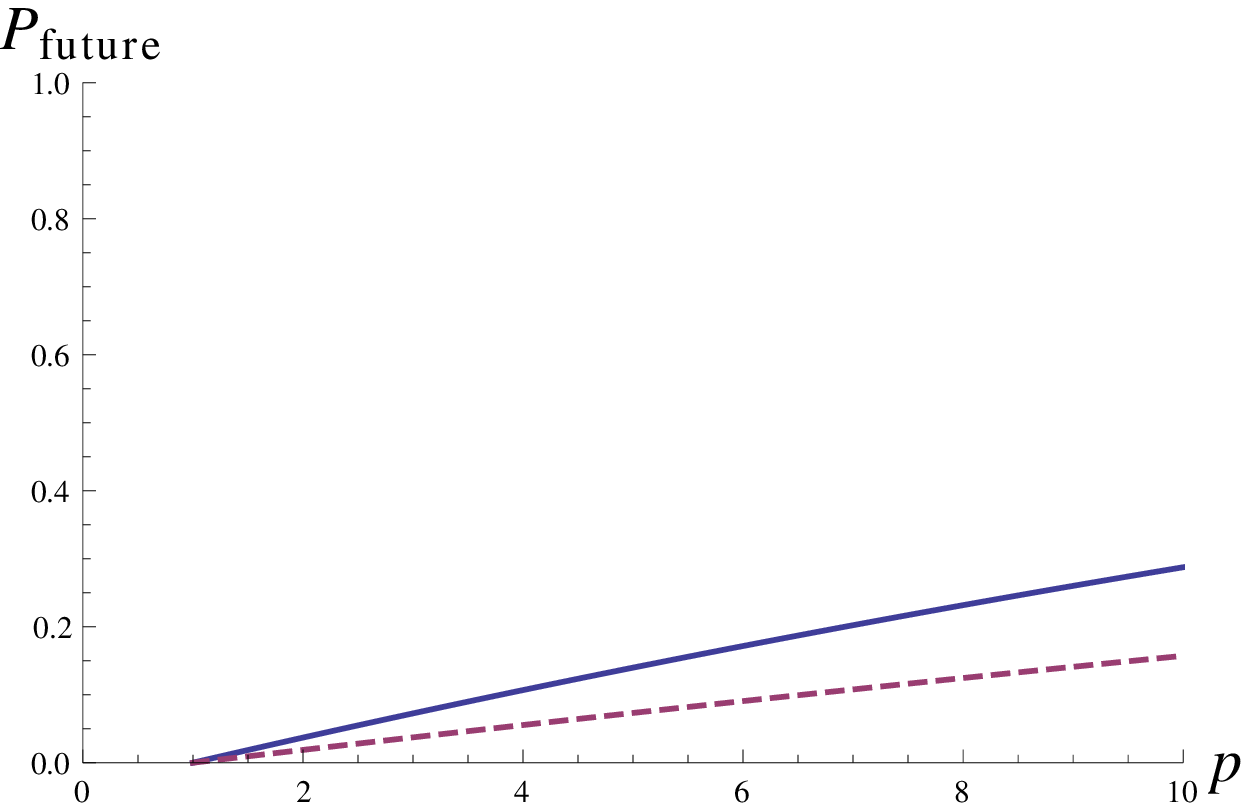}}
  \subfigure[$N_{\rm obs,min} = 40$]{\includegraphics[width=8.5cm]{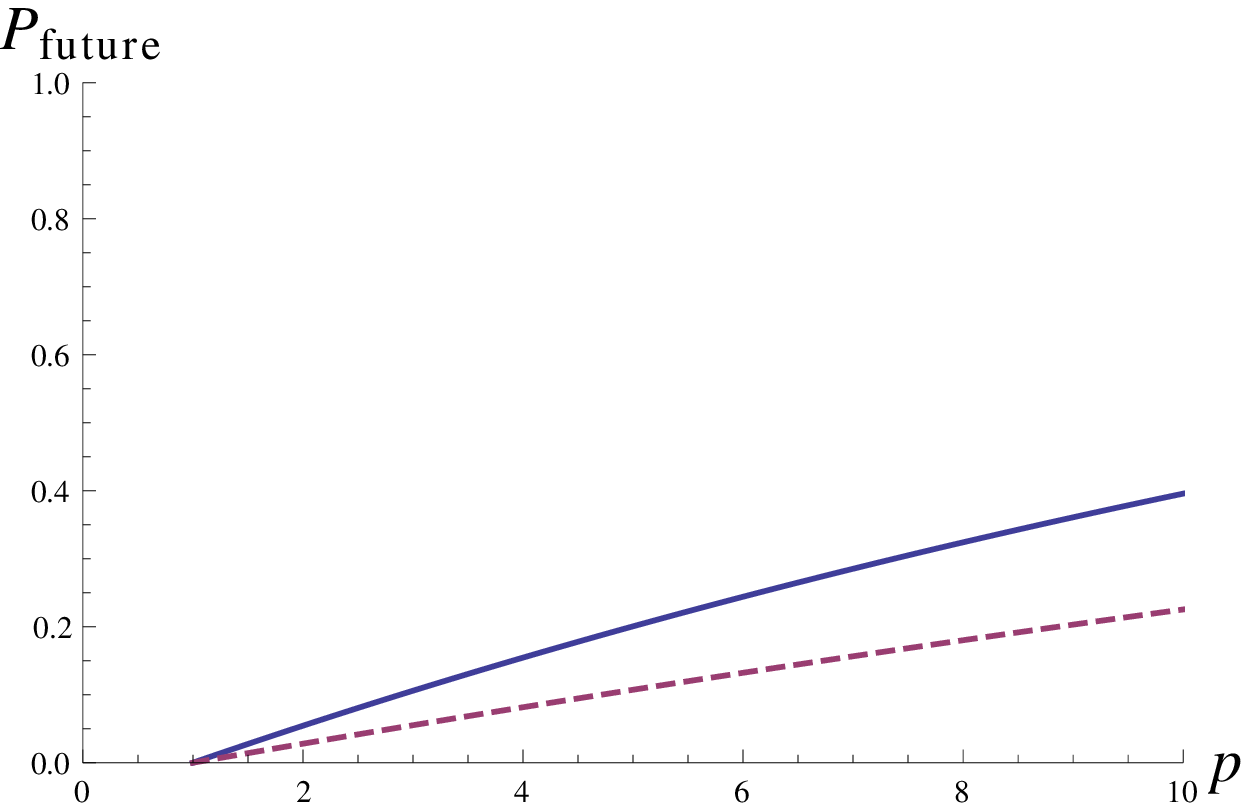}}
\caption{The probabilities of finding nonzero curvature in future 
 measurements at the level of $\Omega_k = 10^{-3}$ (dashed) and 
 $10^{-4}$ (solid) for the multiverse distribution $P(N) \propto 
 1/N^p$.  The probabilities depend on the amount of slow-roll inflation 
 $N_{\rm obs,min}$ corresponding to the maximum curvature allowed 
 by the current observation, $\Omega_{k,{\rm max}} \simeq 0.01$.}
\label{fig:future}
\end{figure}
In Fig.~\ref{fig:p-q-contours}(b), we show contours in the $p$-$q$ 
plane for $P_{\rm future}(10^{-4})$, using $N_{\rm obs,min} = 60$.  In 
Fig.~\ref{fig:future}, we plot the probability for future measurements 
to find $\Omega_k > 10^{-3}$ (dashed) and $10^{-4}$ (solid) as a function 
of $p$, with $q=0$, for $N_{\rm obs,min} = 60$ and $40$.  We find that 
for relatively large $p \simgt \mbox{a few}$, there is a reasonable 
chance that we can observe nonzero curvature larger than $\Omega_k 
\simgt 10^{-4}$.  For $p \simeq 10$, the probability can be as high 
as $\approx 40\%$ for $N_{\rm obs,min} = 40$, which corresponds to 
the case of a (very) low reheating temperature.

In the future, the PLANCK satellite and SDSS will be able to probe 
$\Omega_k$ to the level of $\approx 0.005$~\cite{Eisenstein:1998hr}. 
The planned Subaru surveys also have the potential to reach a $0.3\%$ 
level precision:\ $\sigma(\Omega_k) \approx 0.003$~\cite{Takada:2012}. 
In the longer run, a hypothetical cosmic variance-limited CMB experiment 
together with a measurement of the baryonic acoustic oscillations at 
the precision expected from the Square Kilometer Array will constrain 
curvature with a precision of about $5 \times 10^{-4}$, which can give 
weak evidence for nonzero curvature down to the level of $\Omega_k 
\approx 10^{-3}$~\cite{Vardanyan:2009ft}.  Furthermore, a future 
square kilometer array optimized for $21~{\rm cm}$ tomography could 
improve the sensitivity to about $\sigma(\Omega_k) \approx 2 \times 
10^{-4}$~\cite{Mao:2008ug}, approaching the fundamental limit with 
which one can probe the geometry of the universe given $Q \approx 
10^{-5}$~\cite{Vardanyan:2009ft}.  Therefore, if our own pocket universe 
was indeed created by bubble nucleation in eternally inflating spacetime, 
then there is a reasonable chance (of $O(10\%)$) that we can see nonzero 
negative curvature in future measurements.

\section{Conclusions}
\label{sec:concl}

The eternally inflating multiverse provides a consistent framework for 
explaining coincidences and fine-tuning in our universe.  In particular, 
it provides the leading explanation for the observed accelerating 
expansion of the universe:\ $\Omega_\Lambda \sim \Omega_{\rm matter}$. 
Along similar lines, the framework also provides the possibility that 
the present-day curvature contribution, $\Omega_k$, is not too far 
below the leading contributions to the total energy budget.  Although 
$\Omega_k$ is suppressed exponentially by the deterministic, slow-roll 
inflation that has occurred in our past, $\Omega_k \sim e^{-2N}$, 
there is still a reasonable possibility that $\Omega_k$ is larger 
than $\sim 10^{-4}$, the level we could reach in future observations.

We have studied this possibility, particularly focusing on the question:\ 
``If future observations reveal nonzero curvature, what can we conclude?'' 
We have found that whether an observable signal arises or not depends 
crucially on three issues:\ the cosmic history just before the observable 
inflation, the measure adopted to define probabilities in the eternally 
inflating spacetime, and the properties of the correlation between 
the tunneling and slow-roll parts of the potential.  These strong 
dependencies would allow us to draw some definite conclusions about 
these issues, if nonzero $\Omega_k$ is found in future experiments.

Our conclusions are as follows.  If future measurements reveal positive 
curvature at the level $\Omega_k \simlt -10^{-4}$, then ...
\begin{itemize}
\item
The framework of the eternally inflating multiverse, as currently 
understood, is excluded with high significance.  If no (currently 
unknown) mechanism can be found to explain a positively curved pocket 
universe in an eternally inflating multiverse, then we would have 
to conclude that our universe arose in a different way, e.g.\ directly 
by creation from ``nothing''~\cite{Linde:1995rv}.
\end{itemize}
If future measurements instead reveal negative curvature $\Omega_k 
\simgt 10^{-4}$, then ...
\begin{itemize}
\item
Diffusive (new or chaotic type) eternal inflation is excluded as 
a phenomenon in our immediate past.  In particular, within the context 
of the eternally inflating multiverse (as currently understood), our 
pocket universe must have been born by a bubble nucleation.  In this 
paper we justified this conclusion by examining the evolution of $Q$, 
the density perturbation amplitude, from the end of diffusive eternal 
inflation to the time at which the wave numbers visible in the CMB 
exited the Hubble horizon.  We argued that this evolution required 
more than enough $e$-folds to suppress any trace of curvature.  This 
conclusion is strengthened further by the fact that if the density 
perturbation amplitude was large ($\delta \rho/\rho \sim 1$) 
on the horizon scale at the onset of inflation, then the 
Grishchuk-Zel'dovich effect requires the amount of inflation 
to be large, $N - N_{\rm obs,min} \simgt 6$, completely diluting 
observable curvature effects~\cite{Grishchuk:1978,Kashlinsky:1994rx}. 
The bubble nucleation process avoids this situation by producing, 
without violating causality, a highly homogeneous space that is 
curvature dominated.
\item
Barring the unlikely possibility of a conspiracy between the 
slow-roll volume increase and tunneling rate ($\beta' \simeq -3$; 
see Table~\ref{tab:q}), the probability measure must not reward 
the slow-roll volume increase $e^{3N}$.  Examples of such measures 
include the causal patch measure~\cite{Bousso:2006ev}, the 
scale-factor cutoff measure~\cite{DeSimone:2008bq}, and the 
quantum measure~\cite{Nomura:2011dt}.
\item
The origin of the observed slow-roll inflation---the last $N \approx 
(40~\mbox{--}~60)$ $e$-folds of inflation---must be an accidental 
feature of the potential, selected by anthropic conditions.  In 
particular, it could not be due to a theoretical mechanism that 
ensures the flatness of the potential in the vicinity of our vacuum.
\item
We do not know how to predict the strength or even the sign of possible 
correlations between the tunneling and slow-roll parts of the inflaton 
potential, so we considered all possibilities.  We found that a strong 
negative correlation, one that correlates small $N$ with rapid transitions, 
could have very strong effects which are already excluded by the fact 
that $\Omega_k$ is smaller than is required by anthropic considerations. 
If curvature is observed, then the possibility of strong positive 
correlations (those which favor large $N$) would be ruled out for 
those measures, such as the stationary measure, for which the probability 
of observing a transition depends on the decay rate, and not just the 
branching ratio.  For other measures, the consequence of strong positive 
correlations depends on our estimate of the number of decay channels 
of our parent vacuum that can potentially have a significant branching 
ratio.  If the significant decays are limited to a few hundred nearest 
neighbors in the landscape, then strong positive correlations are 
allowed, and have no perceptible effects on curvature or anything 
else.  On the other hand, if a substantial fraction of the landscape is 
accessible with potentially significant rates, then a strong positive 
correlation would drive a significant increase in $N$, which would be 
ruled out if curvature were observed.
\end{itemize}
If future measurements do not find curvature, $|\Omega_k| \simlt 10^{-4}$, 
then we would not learn much about the structure of the multiverse; in 
particular, it does not support or disfavor the framework.

We also addressed the question of whether the current constraint 
on $\Omega_k \simlt 0.01$ is consistent with the predictions of the 
multiverse picture.  We found that the present constraint is consistent, 
except that for measures that do not reward volume increase, strong 
negative (favoring small $N$) correlations between the slow-roll and 
tunneling part of the potential are ruled out.

In the course of these studies, we were led to consider the 
characteristics of vacuum decay branching ratios, focusing on the 
question of whether decays are typically dominated by a single channel. 
We found that for vacua that are sufficiently long-lived ($S \simgt 
10^3$ if significant decays are limited to several hundred, or $S 
\simgt 10^6$ if decays can access the landscape, where the decay 
rate $\Gamma \sim e^{-S}$), it is plausible that a single channel 
can dominate the decay.

In the next decade or two, we expect to have new data from measurements 
of the CMB, baryonic acoustic oscillations, $21~{\rm cm}$ absorption, 
and so on, which will allow us to probe the curvature of the universe 
down to the level of $\Omega_k \sim 10^{-4}$.  If nonzero $\Omega_k$ 
is found in these measurements, it would reveal another coincidence 
in our universe:\ slow-roll inflation in our past did not last much 
longer than needed to cross the anthropic threshold.  This would 
provide further evidence for the framework of the multiverse.  Moreover, 
it would give us important information about the probability measure, 
the cosmic history just before the observable inflation, and the 
correlations in the inflaton potential function.  In particular, it 
would strongly suggest that the probability measure does not reward 
volume increase, and that we are living in a bubble universe formed 
in an eternally inflating spacetime.

\section*{Acknowledgments}

We thank Asimina Arvanitaki, Savas Dimopoulos, Ben Freivogel, Jenny Guth, 
Larry Guth, and Matthew Kleban for useful discussions.  The work of A.~G. 
was supported in part by the DOE under Contract No.\ DE-FG02-05ER41360. 
The work of Y.~N. was supported in part by the Director, Office of Science, 
Office of High Energy and Nuclear Physics, of the US Department of Energy 
under Contract DE-AC02-05CH11231, and in part by the National Science 
Foundation under Grant No.~PHY-0855653.

\appendix

\section{Volume Increase in the Quantum Measure}
\label{app:quantum}

In the framework of Ref.~\cite{Nomura:2011dt}, the state of the multiverse 
is described in a fixed reference (local Lorentz) frame associated 
with a fixed spatial point $p$.  The Hilbert space corresponding to 
a fixed semi-classical geometry ${\cal M}$ takes the form
\begin{equation}
  {\cal H}_{\cal M} = {\cal H}_{{\cal M}, {\rm bulk}} 
    \otimes {\cal H}_{{\cal M}, {\rm horizon}},
\label{eq:ST-H_M}
\end{equation}
where ${\cal H}_{{\cal M}, {\rm bulk}}$ and ${\cal H}_{{\cal M}, {\rm 
horizon}}$ represent Hilbert space factors associated with the degrees 
of freedom inside and on the stretched apparent horizon $\partial {\cal M}$. 
The entire Hilbert space for dynamical spacetime is then given by the 
direct sum of the Hilbert spaces for different ${\cal M}$'s:
\begin{equation}
  {\cal H} = \bigoplus_{\cal M} {\cal H}_{\cal M}.
\label{eq:ST-H}
\end{equation}
The full Hilbert space for quantum gravity, ${\cal H}_{\rm QG}$, 
also contains the states associated with spacetime singularities, 
${\cal H}_{\rm QG} = {\cal H} \oplus {\cal H}_{\rm sing}$, but the 
states in ${\cal H}_{\rm sing}$ do not play an important role in 
our discussion here.

The multiverse state $\left| \Psi(t) \right>$ is in general a superposition 
of elements in Hilbert space ${\cal H}_{\rm QG}$, and evolves 
deterministically and unitarily in this Hilbert space.  (We take 
the Schr\"{o}dinger picture throughout.)  The probabilities for 
any physical questions can then be given by the (extended) Born 
rule~\cite{Nomura:2011dt}.  For example, one can specify a certain 
``premeasurement'' situation $A_{\rm pre}$ (e.g.\ the configuration 
of an experimental apparatus before measurement) as well as a 
``postmeasurement'' situation $A_{\rm post}$ (e.g.\ those after 
the measurement but without specifying outcome) as $A = \{ A_{\rm pre}, 
A_{\rm post} \}$, and then ask the probability of a particular result 
$B$ (specified, e.g., by a physical configuration of the pointer of 
the apparatus in $A_{\rm post}$) to be obtained.  The relevant probability 
$P(B|A)$ is then
\begin{equation}
  P(B|A) = \frac{\int\!\!\!\int\!dt_1 dt_2 \left< \Psi(0) \right| U(0,t_1)\, 
      {\cal O}_{A_{\rm pre}}\, U(t_1,t_2)\, {\cal O}_{A_{\rm post} \cap B}\, 
      U(t_2,t_1)\, {\cal O}_{A_{\rm pre}}\, U(t_1,0) \left| \Psi(0) \right>}
    {\int\!\!\!\int\!dt_1 dt_2 \left< \Psi(0) \right| U(0,t_1)\, 
      {\cal O}_{A_{\rm pre}}\, U(t_1,t_2)\, {\cal O}_{A_{\rm post}}\, 
      U(t_2,t_1)\, {\cal O}_{A_{\rm pre}}\, U(t_1,0) \left| \Psi(0) \right>}.
\label{eq:prob-final}
\end{equation}
Here, $U(t_1,t_2) = e^{-iH(t_1-t_2)}$ is the time evolution operator 
(for a fixed time parameterization $t$), and ${\cal O}_X$ is the operator 
projecting onto states consistent with condition $X$.  This formula 
can be used to answer questions both regarding global properties of 
the universe and outcomes of particular experiments, providing complete 
unification of the eternally inflating multiverse and the many-worlds 
interpretation of quantum mechanics.

Now, suppose that the probability density for the {\it onset} of 
slow-roll inflation is given by $f(N)$.  To figure out to which class 
the quantum measure belongs, we want to know if the probability density 
of finding an observer {\it at a fixed location with respect to $p$} 
has an extra factor $e^{3N}$ or not (see e.g.~\cite{Larsen:2011mi} for 
relevant discussions).  Since each component of $\left| \Psi(t) \right>$ 
describes the system within the horizon as viewed from $p$, however, 
it is obvious that this extra factor does not exist---i.e., how long 
a state stays in the slow-roll inflation phase does not affect the 
probability defined by Eq.~(\ref{eq:prob-final})---as long as the 
reheating temperature is fixed.  This is because states corresponding 
to different $N$ look identical after the reheating, except for 
quantities that depend on initial conditions at the onset of the 
slow-roll inflation.  And since we are made out of baryons which 
are synthesized after the reheating (i.e.\ whose density does not 
depend on the history before the reheating), the probability density 
of us finding a universe with $N$ $e$-folds of slow-roll inflation is 
simply $f(N)$ in a region where the anthropic factor is unity, $N > 
N_{\rm anthropic}$ (see Section~\ref{subsec:probability}).  This implies 
that the measure belongs to class~(III), according to the classification 
in Section~\ref{subsec:measures}.%
\footnote{Incidentally, if we were made out of relics left over from 
 the era before the inflation, such as the grand unified theory monopole, 
 then the probability of us finding a universe with $N$ $e$-folds 
 of slow-roll inflation would be $f(N) e^{-3N}$ (without taking into 
 account the dynamics for clustering, etc.), since the density of 
 such relics is diluted by the inflation.}

A similar argument implies that the probability does not depend on 
the decay rate of a parent vacuum either.  The quantum measure, therefore, 
gives $q=0$ in Eq.~(\ref{eq:PN-final}); see Table~\ref{tab:q} in 
Section~\ref{subsubsec:dist-N}.

\section{Possibility of Single-Channel Dominance in Multiverse 
 Evolution\protect\footnotemark}
\label{app:single-path}
\footnotetext{We particularly thank Larry Guth for his help with 
this appendix.}

When a metastable vacuum $P_\alpha$ decays, there are generically a very 
large number of decay modes.  One might assume that the decay products 
are dominated by vacua that are nearest neighbors to $P_\alpha$ in the 
landscape, and that the other vacua in the landscape can be neglected. 
In that case, we would expect perhaps several hundred possible decay 
modes.  On the other hand, it is conceivable that a substantial fraction 
of the vacua in the landscape have the possibility of being significant 
decay channels for $P_\alpha$, and then the number of relevant channels 
would be something like 10 to the power of several hundred.  We will 
call the number of relevant decay channels $K$, allowing $K$ to be 
anywhere from several hundred to 10 to the power of several hundred. 
In this appendix we will explore the possibility that this large number 
of decays is dominated by a single channel, finding it much more plausible 
than one might naively guess, especially for long-lived vacua (i.e., 
vacua with decay rates $\Gamma \sim e^{-S}$, where $S \simgt 10^6$ 
for large $K$, or $S \simgt 10^3$ for small $K$).  This issue is relevant 
for Section~\ref{subsubsec:dist}, in discussing the possibilities for a 
multiverse described by Fig.~\ref{fig:model-1} or Fig.~\ref{fig:model-2}, 
and also in Section~\ref{subsubsec:exp}, in estimating the influence 
of nucleation rates on the probability distribution for $N$.

We have no real knowledge of the nucleation rates in the landscape, 
so we will pursue the simple hypothesis that they follow (approximately) 
the normal distribution:
\begin{equation}
  f(S) \approx \frac{1}{\sqrt{2\pi}\,\sigma} 
    e^{-\frac{(S-\bar{S})^2}{2\sigma^2}} ,
\label{eq:distr-S}
\end{equation}
where $\bar{S}$ and $\sigma$ are, respectively, the mean and the standard 
deviation.  We will assume that $\bar{S} \gg \sigma$, so that we can 
ignore the statistically small possibility that the distribution gives 
a negative value for $S$.  Later we will briefly discuss the case where 
$\sigma$ and $\bar{S}$ are comparable.

We now ask:\ for a given $P_\alpha$, what is the typical ratio of the 
fastest decay rate to the next fastest?  Since the number of possible 
decay modes is very large, one might naively think that this ratio is 
close to unity; namely, whatever the fastest rate is, there would likely 
be many other possible decay modes that would have very similar rates. 
This is, however, not obvious because, although the density of the values 
for the decay rates is indeed huge near the peak in the distribution, 
we are interested in the maximum transition rate and the rates that 
are very near the maximum.  These are in the tails of the distribution, 
so there is no guarantee that the naive thinking applies.

To estimate the minimum value of $S$, which we call $S_1$, we define 
the cumulative probability distribution function
\begin{equation}
  \Phi(x) \equiv \int_{-\infty}^x\! f(t)\, dt \, ,
\label{Phi(x)}
\end{equation}
which is the probability that a randomly chosen value of $S$ is less 
than $x$.  We estimate the value of $S_1$ by requiring
\begin{equation}
  \Phi(S_1) = \frac{1}{K}\, ;
\label{eq:S_1-det}
\end{equation}
that is, we imagine drawing $K$ random values $\{S^{(1)},\dots,S^{(K)}\}$ 
from the probability distribution $f(S)$, and insist that the expectation 
value for the number of $S^{(i)}$'s less than $S_1$ is equal to one.

In the region of interest, $(\bar{S}-S_1)/\sigma \gg 1$, the left-hand 
side of Eq.~(\ref{eq:S_1-det}) can be replaced by its asymptotic 
expansion~\cite{Abramowitz}
\begin{equation}
  \frac{1}{\sqrt{2\pi}} 
    e^{-\frac{(\bar{S}-S_1)^2}{2\sigma^2}} 
    \left( \frac{\sigma}{\bar{S}-S_1} - 
    \frac{\sigma^3}{(\bar{S}-S_1)^3} + \cdots \right)=\frac{1}{K}\, ,
\label{eq:Phi-approx}
\end{equation}
giving
\begin{equation}
  S_1 = \bar{S} - \sqrt{2 \ln K}\,\sigma 
    + O\Biggl( \frac{\ln\bigl( \sqrt{\ln K} \bigr)}{\sqrt{\ln K}}\, 
    \sigma \Biggr) \, ,
\label{eq:S_1}
\end{equation}
and
\begin{equation}
  f(S_1) = \frac{\sqrt{2 \ln K}}{K \sigma} 
    + O\Biggl( \frac{\ln\bigl( \sqrt{\ln K} \bigr)}{K \sqrt{\ln K}\, 
    \sigma} \Biggr) \, .
\label{eq:f-S_1}
\end{equation}
Note that $K f(S_1)$ is the density of sample points at $S_1$, so we can 
estimate a typical difference $\varDelta S \equiv S_2 - S_1$, where $S_2$ 
is the second smallest action, as
\begin{equation}
  \varDelta S \approx \frac{\sigma}{\sqrt{2 \ln K}} \, .
\label{eq:varDeltaS}
\end{equation}

The density grows arbitrarily large with $K$, but only as the square 
root of the logarithm!  As we will now see, for reasonable examples 
this is not nearly enough to allow the second fastest decay mode to 
compete with the fastest one.

As an alternative estimate of $\varDelta S$, one could estimate $S_2$ 
directly by setting $\Phi(S_2) = \frac{2}{K}$, which has the effect of 
replacing $K$ by $\frac{1}{2}K$ in Eq.~(\ref{eq:S_1}).  The result for 
$\varDelta S$ is then equal to the result in Eq.~(\ref{eq:varDeltaS}) 
multiplied by $\ln 2$.

It is hard to know what a typical tunneling action is, because various 
calculations have given values over a huge range.  Some of these 
calculations are summarized in Ref.~\cite{DeSimone:2008if}.  For 
example, a calculation of the decay of an uplifted anti--de~Sitter 
vacuum in Ref.~\cite{Ceresole:2006iq} gives an action
\begin{equation}
  S \sim \frac{8 \pi^2 M_{\rm Pl}^2}{m_{3/2}^2} \, ,
\label{eq:Ceresole}
\end{equation}
which the authors estimate as $S \simlt 10^{34}$ using $m_{3/2} \simgt 
10^2~{\rm GeV}$.  Freivogel and Lippert~\cite{Freivogel:2008wm} concluded 
that any vacuum capable of supporting life must decay with an action
\begin{equation}
  S \simlt 10^{40 \pm 20}
\label{eq:Freivogel-Lippert}
\end{equation}
to avoid overproducing Boltzmann Brains, and then showed that 
KKLT~\cite{Kachru:2003aw} vacua decay with actions less than $10^{22}$. 
In Ref.~\cite{Dine:2007er}, however, the authors argue that the vast 
majority of flux vacua with small cosmological constant undergo rapid 
decay, with tunneling actions of order one.

As sample numbers to use here, we consider a transition for which the 
field excursion $\varDelta \varphi$ is of order $M_{\rm Pl}$, while 
the barrier height $\varDelta V_h$ and the energy density difference 
$\varDelta V_{\rm diff}$ are each of $O (M_{\rm unif}^4)$, where 
$M_{\rm unif} \approx 10^{16}~{\rm GeV}$ is the (supersymmetric) 
unification scale.  A small hierarchy between $M_{\rm unif}$ and 
$M_{\rm Pl}$ ensures that metastable minima of the potential are 
long-lived, since the natural size for the action is given by 
$S \approx \frac{27 \pi^2}{2} \varDelta \varphi^4\, \varDelta V_h^2 / 
\varDelta V_{\rm diff}^3$~\cite{Coleman:1977py}.  This estimate gives 
$S \sim O (10^{10})$, and we choose a relatively small $\sigma$, 
$\sigma \sim O (10^8)$.  (Such a narrow distribution of $S$ might 
arise from a structure of the landscape~\cite{ArkaniHamed:2005yv}.) 
We begin by considering $K \sim O(10^{500})$, a number appropriate 
for considering decays to a substantial fraction of the landscape. 
For actions near the peak of the probability distribution, the density 
of sample points per unit of $S$ would then be of order $K/\sigma \sim 
O(10^{492})$, so for every decay channel there would typically be many 
more that would have the same action to hundreds of decimal places. 
Nonetheless, at the tail of the distribution where the fastest 
two decays are to be found, the density of sample points is only 
$\sqrt{2 \ln K}/\sigma$, and $\sqrt{ 2 \ln K}$ is only $\simeq 48$. 
Thus, for our toy numbers the density of sample points in the tail 
is only $\approx 5 \times 10^{-7}$.  This means that the two smallest 
points for $S$ are likely to be separated by $\varDelta S \approx 2 
\times 10^6$, which means that the leading nucleation rate dominates 
over the second place nucleation rate by a factor of $e^{\varDelta S} 
\sim e^{2 \times 10^6}$.  Of course if we used $K$ of order a few 
hundred, the situation would become even more extreme.  For $K \simeq 
200$, for example, $\sqrt{2 \ln K} \approx 3.3$, so the density of 
sample points in the tail is only $\approx 3 \times 10^{-8}$, and 
the leading nucleation rate will dominate over the second place rate 
by a factor of about $e^{\varDelta S} \sim e^{3 \times 10^7}$.

An important caveat of this analysis is the arbitrariness of choosing 
a normal distribution for the values of $S$.  Something resembling 
a normal distribution is plausible, but the actual distribution could 
be very different.  Furthermore, the normal distribution clearly has 
to be modified for cases where it predicts a negative value for $S$. 
For $K \sim O(10^{500})$, Eq.~(\ref{eq:S_1}) implies that the dominating 
value of $S$ is about 48 standard deviations below the mean, so clearly 
the whole approach would break down if $\bar{S} - 48 \sigma$ were not 
positive.  Thus the approach is viable only if $\bar{S} \simgt 50 \sigma$. 
To obtain strong dominance of the leading decay we need $\sigma/50 
\simgt 10^2$, so the argument presented here can lead to the conclusion 
of single-channel dominance only if $\bar{S} \simgt 10^6$. For $K \simeq 
O(200)$ this is less of a problem, because we need only insist that 
$\bar{S} - 3.3 \sigma$ is positive, so a similar argument shows that 
we need only require that $\bar{S} \simgt 10^3$.

If one is interested in parameters for which the normal distribution 
gives negative values of $S$, one could explore the possibility of 
using a probability distribution which is positive by construction. 
A probability distribution that is often used as a model for 
positive-definite quantities is the gamma distribution,
\begin{equation}
  f_\Gamma(S) = \frac{\lambda^{p+1}}{\Gamma(p+1)} S^p e^{- \lambda S} \,
     \theta(S) \, ,
\end{equation}
where $p>0$ and $\lambda > 0$ are parameters to be chosen.  Since we are 
interested only in the low-$S$ tail, we can explore a simpler distribution
\begin{equation}
  f_{\rm simple}(S) = \frac{(p+1) S^p}{S_\rmax^{p+1}}\, \theta(S) \,
     \theta(S_\rmax - S) \, ,
\end{equation}
where $p>0$ and $S_\rmax > 0$ are to be chosen.  Applying the same 
analysis as above, we find that the density of sample points at $S_1$, 
the lowest value of $S$ out of $K$ samples, is given by
\begin{equation}
  K f_{\rm simple}(S_1) \approx \frac{p+1}{S_\rmax}  K^{\frac{1}{p+1}}
     \, .
\end{equation}
If we insist that $K f_{\rm simple}(S_1) \simlt 10^{-2}$ to lead to 
single-channel domination, then with $K \simeq O(200)$ we find that 
$p=2$ allows $S_\rmax \sim 1800$, $p=3$ allows $S_\rmax \sim 1500$, 
while $p=4$ allows $S_\rmax \sim 1450$.  Thus, the new probability 
distribution does not allow us to extend the argument below $\bar{S} 
\sim 10^3$, so we conclude that single-channel dominance is not likely 
to occur for actions this small.  If $K \sim O(10^{500})$, then this 
distribution will never give single-channel dominance.

In summary, these considerations suggest that decays of vacua for 
which the typical action is $\simgt 10^6$ if $K \sim O(10^{500})$, 
or $\simgt 10^3$ if $K \simeq O(200)$, are plausibly dominated by 
a single channel.  This allows for the possibility that the entire 
multiverse is dominated by a single channel.  For example, in the 
scale-factor cutoff measure, the spacetime volume is typically dominated 
by a very slowly decaying, presumably very low energy density vacuum, 
called the dominant vacuum.  An upward tunneling is required to access 
the high-energy part of the multiverse.  In deciding whether the upward 
tunneling is dominated by a single channel, one should keep in mind 
that most of the action appearing in this calculation is associated 
with the initial state, and will apply to all final states; so only 
a small part of the action is relevant for estimating the spread of 
the values for the action.  Nonetheless, it is conceivable that this 
upward tunneling is dominated by a single channel, and that a single 
pathway of subsequent tunnelings dominates the multiverse, as depicted 
in Fig.~\ref{fig:model-2}.  It is also possible, however, that this 
is not the case, and that Fig.~\ref{fig:model-1} is a more accurate 
description of the multiverse.

\end{document}